\documentclass{aa}

\usepackage{graphicx}
\usepackage{txfonts}
\usepackage{natbib}
\usepackage{amssymb}
\usepackage{amsfonts}
\usepackage{amsbsy}
\usepackage{amsmath}

\usepackage{amstext}
\usepackage{pdflscape}
\usepackage{color}
\usepackage{url}
\usepackage{lscape}
\usepackage{longtable}
\usepackage{multirow}

\setcitestyle{plainnat}

\def\Kb~{\rm K-band}
\def\HER~{{\it Herschel} }

\newcommand{\be}{\begin{equation}}
\newcommand{\ee}{\end{equation}}

\newcommand{\etal}{et al..\ }

\newcommand{\Sigmab}{\boldsymbol{\Sigma}}

\newcommand{\Gb}{\boldsymbol{G}}
\newcommand{\Ib}{\boldsymbol{I}}

\newcommand{\phih}{\widehat{\phi}}

\newcommand{\thetah}{\widehat{\theta}}
\newcommand{\Phih}{\widehat{\Phi}}

\newcommand{\Thetah}{\widehat{\Theta}}

\providecommand{\tabularnewline}{\\}

\begin{document}
\title{The bivariate luminosity and mass functions of the local HRS galaxy sample}
\subtitle{The stellar, dust, gas mass functions}

   \author{P. Andreani \inst{1,6} 
 	\and A.~Boselli\inst{2}
	\and L.~Ciesla\inst{2}
	\and R.~Vio\inst{3}
         \and L.~Cortese\inst{4,5} 
        \and V.~Buat\inst{2} 
	\and Y. Miyamoto\inst{6,7} 
}

	\institute{European Southern Observatory, Karl-Schwarzschild-Stra\ss e 2, 85748 Garching, Germany, \email{pandrean@eso.org}   
     \and
        Laboratoire d'Astrophysique de Marseille - LAM, Universit\'e d'Aix-Marseille \& CNRS, UMR7326, 38 rue F. Joliot-Curie, 13388 Marseille Cedex 13, France 
      \and
      Chip Computers Consulting s.r.l., Viale Don L.~Sturzo 82, S.Liberale di Marcon, 30020 Venice, Italy
\and
        International Centre for Radio Astronomy Research, The University of Western Australia, 35 Stirling Hwy, Crawley, WA 6009, Australia 
\and 
ARC Centre of Excellence for All Sky Astrophysics in 3 Dimensions (ASTRO 3D)
 \and
         Nobeyama Radio Observatory, NAOJ, Nobeyama, Minamimaki, Minamisaku, Nagano 384-1305, Japan
 \and
        NAOJ, Osawa, Mitaka, Tokyo 181-8588, Japan
        }
\titlerunning{The stellar, dust, gas mass functions}
\authorrunning{Andreani, Boselli \etal}

\date{Received .............; accepted ................}

\abstract{}{We discuss the results of the relationships between the \Kb~ and stellar mass, far-infrared luminosities, star formation rate, dust and gas masses of nearby galaxies computing the bivariate \Kb~-Luminosity Function (BLF) and bivariate \Kb~-Mass Function (BMF) of the {\textit{Herschel}}~\thanks{\textit{Herschel is an ESA space observatory with science instruments provided by European-led Principal Investigator consortia and with important participation from NASA.}} Reference Survey (HRS), a volume-limited sample with full wavelength coverage.}
{We derive the BLFs and BMFs  from the \Kb~ and stellar mass, far-infrared luminosities, star formation rate, dust and gas masses cumulative distributions using a copula method which is outlined in detail.
The use of the bivariate computed taking into account the upper limits allows us to derive on a more solid statistical ground the relationship between the observed physical quantities.}
{The analysis shows that the behaviour of the morphological (optically selected) subsamples is quite different. {A statistically meaningful result can be obtained over the whole HRS sample only from the relationship between the \Kb~ }{and the stellar mass, while for the remaining physical quantities (dust and gas masses, far-IR luminosity and star formation rate), the analysis is distinct for late-type (LT) and early-type galaxies (ETG). However, the number of ETGs is small to perform a robust statistical analysis, and in most of the case results are discussed only for the LTG subsample.}
The Luminosity and Mass Functions (LFs, MFs) of LTGs are generally dependent on the \Kb~ and the various dependencies are discussed in detail. We are able to derive the corresponding LFs and MFs and compare them with those computed with other samples. Our statistical analysis allows us to characterise the HRS, that, although non homogeneously selected and partially biased towards low IR luminosities, may be considered as representative of the local LT  galaxy population.}{}

\keywords{Galaxies: luminosity function, mass function -- Galaxies: nearby galaxies -- Galaxies: physical process  -- Methods: data analysis -- Methods: statistical}
\maketitle

\section{Introduction}
The way we try to understand galaxy evolution is through the comparison of simulations, both hydrodynamics and semi-analytical, with the physical and statistical properties extracted from the observed galaxy samples. 
One of the extremely useful tool is the abundance matching between a theoretical galactic halo mass function and the observed luminosity and mass functions (LFs, MFs, respectively) of a given population of objects
which provide stringent constrains on the fraction of baryonic mass converted into stars \citep[e.g., see ][]{sha+06}.\hfill\break
Operationally, LFs and MFs are defined as the mean space density of objects per unit luminosity/mass interval \citep[][and references therein]{bin+88,bla+01,bel+03,hil+10,john11}.
A key issue is then to obtain galaxy samples with well defined extracted statistical and physical properties whose selection biases are well under control.
Such samples are difficult to build and require large investment of observing time and of the interpretation of the extracted physical observables.\hfill\break
In the past decades many authors have used local samples selected at various wavelengths to estimate the local LFs and MFs of galaxies and their redshift evolution.
These estimates (and correspondingly the total star formation rates and the local luminosity/mass density) contain some significant uncertainties mainly derived from the lack of either the imaging of large fields, or the required multi-wavelength homogeneous coverage and complete redshift information.\hfill\break
Any astronomical sample is affected by selection effects and systematic biases, therefore any statistically meaningful inference of the LF and MF needs a careful analysis of these issues.
Even with the best data sets the accurate construction of the LF/MF remains a tricky pursuit, since the presence of observational
selection effects due to e.g. detection thresholds in apparent magnitude, colour, surface brightness or some combination thereof can
make any given galaxy survey incomplete and thus introduce biases in the LF/MF estimates.
This is particularly critical by investigating the LFs at other wavelengths than that used as primary selection criterium of the sample. 
In this latter case the use of the bivariate LFs (BLFs, and in the case of the mass functions, BMFs), if the statistical assumptions are correctly defined, may provide a powerful method of studying the LFs at wavelengths different from the selection one.\hfill\break
The \HER~ Reference Survey (HRS) is a \HER~ guaranteed time key project, performing photometric observations with the SPIRE cameras towards HRS galaxies \citep{bos+10}.
The HRS is a volume-limited sample (i.e., 15$<D<$25 {\rm Mpc}) including late-type galaxies (LTGs) (Sa and later) with 2MASS \Kb~ magnitude $\leq$12 {\rm mag} and early-type galaxies (ETGs) (S0a and earlier) with $\leq$ 8.7 {\rm mag}. 
The survey selection criteria (magnitude- and volume-limited, see \S ~\ref{sample}), size and multiwavelength coverage (from UV to radio wavelengths both in spectroscopy and photometry) together with the \HER~ results in the far-IR, sensitive to dust mass down to $10^4 M_\odot$,
have shown that the HRS can be considered as a 'reference' sample to carry out statistical analysis in the local Universe \citep{bos+10}. \hfill\break
Its use as a reference sample has been key to compare the predicted scaling relations (dust-to-stellar mass ratios and gas fraction) \citep{mck+16,dav+17} 
to provide additional constraints on feedback mechanisms and other physical processes of galaxy formation in cosmological simulations \citep[i.e.][]{lag+16}.\hfill\break
With the above in mind, we use in this paper the HRS sample to investigate the BLFs and the BMFs derived in the various frequency bands. 
{The HRS sample is \Kb~ }{selected and a {\it direct} derivation of the LF can be carried out only at this wavelength \citep{bos+10,PaperI}, although with a small statistical significance because of the small number of sources. At wavelengths different from that of the selection there is no way to unbiasedly derive a LF or a MF
\citep[for a complete discussion see][]{john11}.

In this paper we exploit the knowledge of the \Kb~ LF, which is well established \citep{col+01,koc+01} and use the BLF as a statistical tool in the presence of upper limits which
provides results that lie on more solid statistical ground than any other simpler tool, i.e. a linear regression test.
The HRS is relative small to this aim and limited in statistics but it is the only sample with a complete and accurate multiwavelength coverage.\hfill\break
\citet{PaperI} have already determined the BLF of the HRS sample but restricted to the {\it monochromatic} cases: \Kb~ - 250$\mu$m, 
 \Kb~ - 350$\mu$m,  \Kb~ - 500$\mu$m.
Meanwhile with the collection and the analysis of a larger multiwavelength dataset \citep{bos+13,bos+14a,bos+15,cies+14,cor+12a} we are able to extend our analysis. We make use of the total IR and the H$\alpha$ luminosities, the stellar, dust and gas masses to derive the
bivariate functions with respect to the \Kb~ luminosity, which is the band at which the sample is complete
\citep{PaperI}. Our analysis is statistically clean because of the accuracy of the employed statistical method and of the use of the whole data sets including the upper limits on the observed fluxes (and therefore on the derived quantitied from these fluxes).\hfill\break
We compare the derived functions to the ones computed for other local samples, to understand how the selection criteria affect the outcomes. These latter must be taken into account when comparing simulations with observations.
The distribution functions in H$\alpha$, ${\rm HI}$ and ${\rm H_2}$ have been already studied in dedicated works \citep{bos+14b,bos+15} and their properties and differences with respect to other local samples discussed in those papers.\hfill\break
The paper is organised as follows. The sample is briefly described in section~\ref{sample}. The mathematical tools and the method used to compute the luminosity functions are described in section~\ref{tools}. Results are described in section~\ref{results} and discussed in section~\ref{discussion}. Conclusions are summarised in section~\ref{conclusions}.

\section{The data}
\label{sample}

The HRS is a volume-limited sample (i.e., 15$<D<$25 {\rm Mpc}) including late-type galaxies (LTGs) (Sa and later) with 2MASS \Kb~ magnitude $\leq$12 {\rm mag} and early-type galaxies (ETGs) (S0a and earlier) with $\leq$ 8.7 {\rm mag}. Additional selection criteria are high Galactic latitude ($b >+55^\circ$)
and low Galactic extinction (AB$<$0.2 {\rm mag}, \citep{sch+98}).
The sample includes 322 galaxies (260 LTGs and 62 ETGs), and the total volume over an area of 3649 sq.deg. is 4539 Mpc$^3$.
The selection criteria are fully described in \citet{bos+10}.\hfill\break
The multiwavelength data used in this work have been taken from \citet{bos+11}, \citet{bos+13}, \citet{bos+15}, \citet{cies+12}, \citet{cor+12a}, \citet{cor+14}.
Morphological types and distances are taken from \citet{cor+12a}.

This huge data set has been extensively used to derive and discuss the main physical properties of this sample.
In this work we make use of the stellar masses,
determined from $L_i$ and ${g-i}$ \citep{cor+12a} following the prescription of \citet{zib+09} based on the $i$-band luminosity and $g-i$ mass-to-light ratio. 
For galaxies without SDSS $g$ and $i$-band data, stellar masses have been computed
using the prescription of \citet{bos+09} based on the H-band luminosity and $B-H$ mass-to-light ratio.
The stellar mass range covered by this sample is $8<log(M_{\rm star}/M_\odot)<12$. 
The total FIR Luminosity and the dust masses are derived from the SED fit \citep{cies+14}, the gas masses (from CO and HI observations, and the molecular mass given in Table~\ref{tabledata} {determined using a luminosity dependent value as explained in detail in \citet{bos+14a,bos+14b}.
A constant $X_{\rm CO}$ factor, usually employed to convert the CO luminosity to molecular gas mass, underestimates the
molecular content at stellar masses below $10^{10}M\odot$.

SFR are taken from \citet{bos+15} and are the average of the values derived from H$\alpha$ luminosities (from \citet{bos+15}) corrected using the Balmer decrement (from \citet{bos+13}) or the far-IR emission at 24 $\mu$m (from \citet{ben+12} and \citet{cies+14}), from GALEX FUV luminosities (from \citet{cor+12b}) still corrected for dust attenuation using the far-IR emission at 24 $\mu$m,
and from 20 cm radio luminosities (collected mainly from the NVSS \citep[see,][]{bos+15}). The choice of using different tracers has been done to minimise 
the observational errors and the uncertainty on the dust attenuation correction and to have at least one measure for each galaxies (not all data are available for the whole sample). Details on the adopted calibrations and corrections can be found in \citet{bos+15}.

Because the stellar masses have been computed using a Chabrier IMF, to use consistent values for stellar masses and SFR, we convert the SFR, which have been derived using a Salpeter IMF, to values compatible with a Chabrier IMF dividing the first SFR by 1.58.\hfill\break
The sample has a very limited luminosity coverage, the maximum observed luminosity at 250$\mu$m is $10^9$L$_\odot$, and it contains the Virgo cluster which might introduce two biases. (1) Morphology segregation effect \citep{dres80}: clusters are dominated by ETGs compared to the field. 
The HRS contains a higher fraction of ETGs than one would normally find in a "blindly generated" sample, as for instance the H-ATLAS \citep{vac+10} and the HerMES survey \citep{mar+16}, where the fraction of cluster galaxies is only a few percent.
(2) The LTGs in clusters are different compared to those in the field for multiple reasons.
For instance they have a reduced star formation and therefore a reduced far-infrared emission
because they are poorer in gas \citep{BG06,BG14,bos+14c,bos+14d}. \citet{cor+10,cor+12b} have shown that the HRS LTGs in the VIRGO cluster have truncated dust discs and lower dust masses.
This might introduce a non homogenous \Kb~ distribution for LTGs because of the presence of two types of LTs: cluster and field galaxies.
However, as already shown in \citet{bos+10} and in \citet{PaperI}, the \Kb~ LF computed on the HRS sample agrees within the errorbars with the LFs computed on the parent sample (the 2MASS) \citep{koc+01,col+01} despite the limited range in luminosities spanned from the HRS.\hfill\break
Additional information about this sample may be found in \citet{bos+10,bos+15} and \citet{cor+12a}.

\onecolumn
\begin{longtable}{ccccccccc}
\caption{Logarithmic values of the luminosities, masses and the star formation rates of the HRS. \label{tabledata}}\\
\hline\hline
\\
\multicolumn{1}{c}{HRS} &
\multicolumn{1}{c}{$L(\rm H\alpha)$} &   
\multicolumn{1}{c}{$L({\rm K}$)}      & 
\multicolumn{1}{c}{$L({\rm IR}$)}     &
\multicolumn{1}{c}{$M_{\rm dust}$}    & 
\multicolumn{1}{c}{$M_{\rm star}$}  & 
\multicolumn{1}{c}{$M_{\rm HI}$}       &
\multicolumn{1}{c}{$M_{\rm H2}$}  &    
\multicolumn{1}{c}{SFR}   \\
\multicolumn{1}{c}{} &
\multicolumn{1}{c}{($erg s^{-1}$)} &      
\multicolumn{1}{c}{($L_\odot$)} &    
\multicolumn{1}{c}{($L_\odot$) } &     
\multicolumn{1}{c}{($M_\odot$)} &    
\multicolumn{1}{c}{($M_\odot$)}  & 
\multicolumn{1}{c}{($M_\odot$)} &   
\multicolumn{1}{c}{($M_\odot$)}      & 
\multicolumn{1}{c}{($M_\odot$/yr)}     \\
\multicolumn{1}{c}{(1)} &
\multicolumn{1}{c}{(2)}&      
\multicolumn{1}{c}{(3)} &    
\multicolumn{1}{c}{(4)} &     
\multicolumn{1}{c}{(5)} &    
\multicolumn{1}{c}{(6)}  & 
\multicolumn{1}{c}{(7)} &   
\multicolumn{1}{c}{(8)}      & 
\multicolumn{1}{c}{(9)}      
\\
\hline
  \\
\endhead 
 \\ 
 \hline
  \multicolumn{9}{c}{{Continued on next page\ldots}} \\
\endfoot

  \\ 
\hline \hline
\endlastfoot 
 1 & 39.17 &  9.83 & 8.48 & 5.93 & 9.10 & ---    & --- &  0.05010 \\
 2 & 39.84 &  9.96 & 9.08 & 6.23 & 9.11 & 8.42 & --- &  0.19272 \\
 3 & 38.39 & 11.08 & 8.56 & 5.87 & 10.47 & 8.65 & 8.05 &  0.00379 \\
 4 & 38.55 & 11.5  & 9.9 & 7.41 & 10.63 & 8.98 & 9.17 &  0.75207 \\
 5 & 37.49 & 10.4  & 9.2 & 6.58 & 9.7 & 8.2 & --- &  0.13443 \\
 6 &  ---     & 9.9 & 8.21 & 6.92 & 9.08 & 8.89 & --- &  0.02582 \\
 7 &  ---     & 11.36 & 9.2 & 6.13 & 10.64 & 7.63 & 7.86 &  0.00131  \\
 8 & 38.18 & 11.04 & 9.19 & 7.46 & 10.16 & 9.66 & 8.39 &  0.17824  \\
 9 & 38.91 & 10.97 & 9.06 & 6.5 & 10.24 & 8.65 & 8.57 &  0.10294 \\
 10 & 38.98 & 10.06 & 8.83 & 6.19 & 9.2 & 8.76 & --- &  0.15716 \\
11 & 38.86 & 10.56 & 9.38 & 6.9 & 9.71 & 8.79 & 8.43 &  0.38172\\
 12 & 39.01 & 9.69 & 8.42 & 5.72 & 8.85 & 8.23 & --- &  0.06024   \\
13 & 39.46 & 11.44 & 10.16 & 7.54 & 10.54 & 9.32 & 8.80 &  1.86583  \\
 14 &  ---  & 11.08 & 8.61 & 5.45 & 10.2 & 7.79 & 7.78 &  0.0005   \\
15 & 38.69 & 11.3 & 9.8 & 7.66 & 10.39 & 9.97 & 8.94 &  0.99652  \\
16 & 39.41 & 10.84 & 9.44 & 7.21 & 9.97 & 9.14 & 8.50 &  0.5234   \\
 17 & 38.58 & 10.75 & 9.67 & 7.15 & 9.87 & 9.42 & 8.93 &  0.80073  \\
 18 & 38.8 & 10.71 & 9.11 & 6.54 & 9.93 & 8.54 & --- &  0.1366   \\
 19 & 38.72 & 10.65 & 9.48 & 6.94 & 9.79 & 9.32 & --- &  0.60928  \\
 20 & 39.09 & 10.77 & 9.98 & 7.21 & 9.92 & 9.47 & 8.14 &  1.90571   \\
21 & 38.38 & 9.75 & 8.31 & 6.35 & 8.99 & 8.28 &  --- &  0.01456   \\
22 & 37.88 & 11.38 & 8.32 & 5.98 & 10.88 & 8.12 & 7.65 & 0.00452   \\
 23 & 39.45 & 11.03 & 9.96 & 7.31 & 10.2 & 9.17 & 8.16 &  1.03844   \\
 24 & 38.67 & 11.17 & 9.91 & 7.49 & 10.31 & 10.04 & 8.94 &  1.2722   \\
25 & 39.13 & 10.92 & 10.03 & 7.04 & 10.0 & 9.24 & 8.72 &  1.41582   \\
26 &  ---  & 9.96 & 8.69 & 6.24 & 9.15 & 8.56 &  --- &   0.06486  \\
 27 & 39.63 & 10.28 & 9.47 & 6.49 & 9.44 & 8.62 & 9.06 &  0.57398   \\
  28 & 38.75 & 10.39 & 9.3 & 6.72 & 9.48 & 8.85 & --- &   0.34901   \\
 29 &  ---  & 10.07 & 8.91 & 6.56 & 9.27 & 8.54 &  --- &   0.07496   \\
  30 & 38.85 & 10.18 & 8.99 & 6.67 & 9.4 & 9.12 &  --- &  0.19726   \\
 31 & 38.03 & 11.08 & 9.78 & 6.97 & 9.94 & 9.77 & 8.32 &  1.07332   \\
  32 &  ---  & 10.49 & 7.4 & 5.97 & 9.75 & 7.92 & 7.58 &  0.00015   \\
  33 & 38.76 & 10.83 & 9.45 & 7.01 & 9.96 & 9.41 & 8.61 & 0.50899   \\
 34 & 37.37 & 10.45 & 9.27 & 7.11 & 9.76 & 9.14 & 8.64 &  0.21611   \\
 35 & 38.99  & 10.64 & 9.34 & 6.71 & 9.81  & ---  & 7.69  & 0.00530  \\
  36 & 39.17 & 11.35 & 10.35 & 7.44 & 10.49 & 8.74 & 9.25 &  2.93756  \\
  37 & 38.99 & 10.64 & 9.34 & 6.71 & 9.81 & 8.84 & 8.62 &  0.41618   \\
 38 &  ---  & 10.28 & 8.99 & 6.59 & 9.4 & 9.0 & 8.30 &  0.21554   \\
  39 &  ---  & 10.13 & 8.75 & 6.7 & 9.36 & 9.13 & --- &  0.08668  \\
 40 & 38.7 & 10.49 & 9.48 & 6.67 & 9.61 & 9.02 & --- &  0.52878  \\
41 & 37.82 & 10.24 & 8.71 & 6.55 & 9.45 & 8.54 & --- &  0.0455   \\
  42 & 38.66 & 11.07 & 9.71 & 7.31 & 10.15 & 9.39 & 8.83 & 0.92682   \\
43 &  ---  & 11.08 & 8.15 & 6.06 & 10.59 & 7.09 & 7.81 &  0.00092  \\
  44 & 39.52 & 9.88 & 8.94 & 6.08 & 9.01 & 8.37 & --- &  0.17407   \\
 45 &  ---  & 11.19 & 9.1 & 6.88 & 10.48 & 8.89 & 8.17 &  0.05492   \\
  46 &  ---  & 11.34 & 9.48 & 6.58 & 10.59 & 9.02 & 8.17 &  0.09389   \\
  47 & 38.57 & 10.49 & 9.22 & 6.79 & 9.56 & 9.32 & --- & 0.45388   \\
  48 & 38.88 & 11.21 & 10.09 & 7.5 & 10.31 & 9.51 & 9.27 &  2.10298   \\
  49 &  ---  & 11.39 & 8.29 & 6.23 & 10.69 & 8.27 & 7.86 &  0.00081   \\
50 & 39.79 & 11.05 & 10.11 & 7.18 & 10.17 & 9.23 & 9.02 & 1.56473   \\
51 & 38.88 & 10.38 & 9.3 & 6.8 & 9.53 & 9.31 &  ---  &  0.36642  \\
  52 &  ---  & 10.44 & 8.69 & 5.98 & 9.69 & 9.35 & --- &  0.09075   \\
  53 & 38.81 & 10.66 & 9.46 & 6.94 & 9.81 & 9.39 & 8.65 & 0.41137   \\
  54 & 38.76 & 10.83 & 9.07 & 6.65 & 10.09 & 9.41 & 8.14 &  0.1882   \\
 55 & 38.91 & 10.69 & 9.51 & 7.04 & 9.82 & 9.27 & 8.72 &  0.51186  \\
 56 & 39.48 & 11.27 & 10.06 & 7.84 & 10.57 & 9.06 & 9.15 &  0.98898  \\
  57 & 39.05 & 11.14 & 9.71 & 7.24 & 10.22 & 8.91 & 8.82 &  0.79531  \\
  58 & 38.88 & 10.1 & 8.79 & 6.32 & 9.27 & 8.29 &  ---  &  0.13965  \\
  59 & 38.78 & 11.0 & 9.51 & 7.24 & 10.26 & 9.24 & 8.65 &  0.34514  \\
  60 & 39.15 & 10.95 & 9.47 & 6.84 & 10.14 & 9.47 & 8.72 &  0.25265  \\
 61 &  ---  & 9.83 & 8.44 & 6.2 & 8.95 & 9.01 &  ---  &  0.08972  \\
 62 & 38.7 & 10.47 & 9.24 & 7.04 & 9.59 & 9.65 &  ---  &   0.43629  \\
  63 & 38.35 & 11.22 & 9.54 & 7.46 & 10.3 & 9.34 & 8.67 &  0.46393  \\
  64 &  ---  & 10.19 & 8.71 & 6.59 & 9.38 & 8.76 &  ---  &   0.06966  \\
  65 &  ---  & 10.29 & 9.03 & 6.56 & 9.46 & 9.27 &  ---  &   0.27295  \\
 66 & 39.38 & 11.04 & 10.12 & 7.3 & 10.15 & 9.40 & 9.07 &  1.81949  \\
  67 &  ---  & 10.14 & 8.76 & 6.57 & 9.27 & 8.96 &  ---  &  0.15135  \\
 68 &  ---  & 9.97 & 8.97 & 5.87 & 9.16 & 8.01 &  ---  &   0.11025  \\
  69 & 38.74 & 11.4 & 9.0 & 7.14 & 10.65 & 9.49 & 8.05 &  0.07897  \\
  70 & 38.92 & 10.18 & 9.11 & 6.55 & 9.28 & 8.92 &  ---  & 0.33333  \\
  71 &  ---  & 11.43 & 8.94 & 7.22 & 10.55 & ---   &  8.09 & 0.001604     \\
 72 &  ---  & 10.2 & 9.15 & 6.51 & 9.27 & 9.23 &  ---  &  0.36979   \\
 73 & 38.09 & 11.44 & 10.01 & 7.77 & 10.73 & 9.40 & 9.32 &  0.70618   \\
 74 & 39.14 & 10.81 & 9.78 & 6.92 & 9.9 & 9.12 & 8.85 &  0.99546   \\
  75 &  ---  & 9.87 & 8.11 & 6.39 & 9.17 & 8.65 &  ---  &   0.00868  \\
  76 & 38.37 & 9.78 & 8.43 & 6.07 & 8.96 & 8.63 &  ---  &  0.11008   \\
  77 & 39.4 & 11.65 & 10.5 & 7.96 & 10.82 & 9.82 & 9.29 &  3.83381   \\
  78 & 38.96 & 10.3 & 9.0 & 6.61 & 9.46 & 9.23 & 8.74 &  0.21552   \\
  79 & 38.1 & 10.04 & 8.93 & 6.31 & 9.15 & 9.12 &  ---  &   0.22143   \\
  80 & 38.89 & 10.5 & 8.69 & 6.61 & 9.47 & 8.53 &  ---  &   0.07175   \\
  81 & 38.36 & 10.93 & 9.86 & 7.12 & 10.18 & 9.08 & 9.15 &  0.62496   \\
  82 &  ---  & 9.87 & 8.56 & 5.83 & 8.79 & 8.07 &  ---  &   0.05461   \\
  83 &  ---  & 9.8 & 8.4 & 5.82 & 8.93 & 8.09 &  ---  &   0.05923   \\
  84 & 39.49 & 10.43 & 8.97 & 6.42 & 9.76 & 8.52 & 8.39 &  0.14092   \\
  85 & 38.37 & 11.28 & 9.92 & 7.32 & 10.4 & 9.38 & 9.04 &  0.86625   \\
  86 & 38.19 & 10.27 & 9.45 & 7.21 & 9.75 & 9.54 & 8.67 &  0.65953   \\
  87 &  ---  & 10.92 & 8.72 & 6.25 & 10.31 & 7.34 &  ---  &   0.01464   \\
  88 & 38.36 & 11.21 & 9.8 & 7.44 & 10.03 & 9.58 & 8.81 &  1.03332   \\
  89 & 38.52 & 10.72 & 9.69 & 7.55 & 10.02 & 9.63 & 8.41 &  0.93302   \\
  90 &  ---  & 11.17 & 8.29 & 6.49 & 10.51 & 7.80 & 7.85 &  0.00117   \\
  91 & 38.61 & 11.71 & 10.17 & 7.88 & 10.93 & 9.70 & 9.07 &  0.99281   \\
  92 & 38.02 & 10.13 & 8.93 & 6.57 & 9.4 & 8.85 &  ---  &  0.09066   \\
  93 &  ---  & 11.34 & 9.0 & 6.54 & 10.86 & 9.44 & 7.98 &  0.00482   \\
  94 & 37.81 & 10.58 & 9.18 & 7.26 & 9.86 & 9.44 & 8.59 &  0.19435   \\
  95 & 38.95 & 10.59 & 9.55 & 6.64 & 9.92 & 8.65 & 8.46 &  0.29365   \\
 96 & 39.02 & 11.18 & 9.96 & 7.35 & 10.36 & 8.98 & 8.99 &  0.92468   \\
  97 & 38.62 & 11.79 & 9.79 & 8.06 & 11.04 & 9.33 & 8.84 &  0.007   \\
  98 & 38.33 & 10.46 & 9.22 & 7.03 & 9.72 & 9.10 & 8.29 &  0.1619   \\
  99 & 38.8 & 9.88 & 8.69 & 6.03 & 9.19 & 7.85 &  ---  &   0.07007   \\
 100 & 38.81 & 10.92 & 9.67 & 7.08 & 10.16 & 8.37 & 9.00 &  0.29616   \\
 101 &  ---  & 11.19 & 8.22 & 5.65 & 10.59 & 7.34 & 7.62 &  0.00095   \\
 102 & 39.1 & 11.57 & 10.69 & 8.05 & 10.81 & 9.72 & 9.73 &  7.20117   \\
 103 & 37.88 & 11.23 & 8.85 & 6.61 & 10.76 & 7.73 & 8.76 &  0.00154   \\
 104 & 36.54 &  9.66  & 7.58  & 5.43   & 8.80 &  --- & ---  & 0.002261    \\
 105 &  ---  & 11.03 & 8.17 & 6.15 & 10.45 & 8.77 & 7.66 & 0.00216   \\
 106 & 38.42 & 10.44 & 9.18 & 6.8 & 9.7 & 8.80 &  ---  &  0.22798   \\
 107 & 37.45 & 9.97 & 8.7 & 6.51 & 9.38 & 8.44 &  ---  &  0.04226   \\
 108 & 38.39 & 10.14 & 8.91 & 6.29 & 9.45 & 8.14 &  ---  &   0.06854   \\
 109 & 37.28 & 10.42 & 9.1 & 6.78 & 9.72 & 9.20 &  ---  &  0.10702   \\
 110 & 38.41 & 10.51 & 9.48 & 6.92 & 9.71 & 9.31 & 8.06 &  0.57213   \\
 111 & 38.91 & 11.05 & 9.84 & 7.4 & 10.32 & 9.09 & 9.07 &  0.56884   \\
 112 & 38.25 & 10.86 & 8.85 & 6.35 & 10.21 & 7.86 & 8.18 &  0.01432   \\
 113 & 38.31 & 11.25 & 9.92 & 7.73 & 10.6 & 9.35 & 9.14 &  0.38418   \\
 114 & 39.05 & 11.62 & 10.69 & 7.94 & 10.86 & 9.76 & 9.63 &  7.68282   \\
 115 & 38.08 & 10.04 & 8.67 & 6.05 & 9.41 & 7.74 &  ---  &   0.0156   \\
 116 &  ---  & 10.43 & 8.37 & 6.2 & 9.83 & 7.49 &  ---  &   0.00079  \\
 117 & 37.97 & 11.17 & 9.56 & 7.35 & 10.53 & 8.22 & 8.97 &  0.01134   \\
 118 & 38.51 & 9.92 & 8.71 & 6.31 & 9.17 & 9.16 &  ---  &   0.22161   \\
 119 & 37.29 & 10.8 & 9.47 & 6.85 & 10.14 & 8.12 & 9.1 &  0.00224   \\
 120 & 37.79 & 11.01 & 9.32 & 6.98 & 10.37 & 8.05 & 8.75 &  0.00273   \\
 121 & 38.29 & 10.94 & 9.7 & 7.28 & 10.29 & 9.08 & 8.45 &  0.28189   \\
 122 & 38.71 & 11.8 & 10.61 & 8.16 & 11.01 & 9.52 & 9.55 & 4.89071  \\
  123 & 38.23 & 11.0 & 8.99 & 6.69 & 10.31 & 8.79 & 7.78 & 0.05714   \\
  124 & 37.37 & 10.6 & 9.21 & 7.08 & 9.82 & 8.67 & 8.72 &  0.1815   \\
  125 &  ---  & 11.25 & 8.5 & 6.68 & 10.76 & 7.88 & 7.92 &  0.00326   \\
  126 &  ---  & 11.14 & 8.01 & 5.81 & 10.41 & 7.30 & 7.91 &  0.00082   \\
  127 & 38.59 & 11.06 & 9.57 & 7.17 & 10.4 & 8.84 & 9.06 &  0.17725   \\
  128 & 38.24 & 10.25 & 8.87 & 6.81 & 9.59 & 8.44 &  ---  &   0.09821   \\
  129 &  ---  & 11.27 & 8.7 & 5.55 & 10.29 & 7.42 & 7.78 &  0.00063   \\
  130 & 37.97 & 10.33 & 8.99 & 6.56 & 9.56 & 8.58 & 8.31 &  0.08867  \\
  131 & 38.22 & 10.24 & 8.85 & 6.62 & 9.53 & 8.35 &  ---  &  0.08434   \\
  132 & 38.33 & 10.18 & 9.11 & 6.42 & 9.47 & 8.94 &  ---  &   0.32065   \\
  133 & 38.24 & 10.52 & 8.88 & 7.01 & 9.6 & 9.23 & 8.39 &  0.11477   \\
  134 & 37.91 & 10.84 & 9.18 & 6.77 & 10.12 & 8.06 &  ---  &  0.00574   \\
  135 &  ---  & 12.04 & 8.97 & 6.72 & 11.54 & 8.13 & 8.31 &  0.01339   \\
  136 &  ---  & 10.92 & 9.32 & 6.78 & 10.29 & 7.95 & 8.53 &  0.00113   \\
  137 &  ---  & 11.3 & 8.24 & 6.39 & 10.78 & 7.46 & 7.97 &  0.00085   \\
  138 & 37.64 & 11.82 & 8.77 & 5.58 & 11.15 & 7.50 & 8.00 &  0.00118   \\
  139 & 38.17 & 10.21 & 9.1 & 6.66 & 9.46 & 8.89 &  ---  &   0.32062   \\
  140 & 38.61 & 11.0 & 9.05 & 7.09 & 10.37 & 8.90 & 8.28 &  0.08475   \\
  141 & 38.03 & 11.33 & 9.54 & 7.49 & 10.6 & 8.51 & 8.66 &  0.28389   \\
  142 & 38.87 & 10.62 & 9.81 & 6.73 & 9.76 & 9.48 & 8.56 &  1.08645   \\
  143 & 38.13 & 10.84 & 9.73 & 7.34 & 10.02 & 9.53 &  ---  &  0.65339   \\
  144 & 38.47 & 11.22 & 10.06 & 7.17 & 10.42 & 8.71 & 8.64 & 0.88967   \\
  145 & 38.62 & 10.44 & 9.27 & 7.01 & 9.69 & 8.96 & 8.75 &  0.3342   \\
  146 & 38.63 & 10.45 & 9.27 & 6.9 & 9.7 & 8.76 &  ---  & 0.16605   \\
  147 & 37.51 & 10.98 & 9.02 & 7.0 & 9.62 & 8.72 &  ---  &   0.04507   \\
  148 & 38.48 & 10.27 & 9.27 & 7.09 & 9.53 & 9.05 & 8.74 & 0.37708   \\
  149 & 39.26 & 11.03 & 9.89 & 7.47 & 10.36 & 8.72 & 9.23 & 0.57647   \\
  150 & 38.02 & 11.95 & 8.81 & 6.78 & 11.19 & 7.95 & 7.52 &  0.00088   \\
  151 & 38.4 & 10.49 & 9.2 & 6.77 & 9.79 & 8.33 & 8.49 &  0.1302   \\
  152 & 38.99 & 10.48 & 9.53 & 6.48 & 9.71 & 8.40 & 8.59 &  0.52501   \\
  153 & 38.52 & 10.33 & 9.2 & 6.76 & 9.6 & 8.46 & 8.44 &  0.22311   \\
  154 & 38.06 & 10.58 & 9.23 & 7.17 & 9.75 & 9.42 & 8.26 &  0.32224   \\
  155 &  ---  & 11.36 & 8.4 & 6.32 & 10.76 & 7.74 & 7.87 &  0.00145   \\
  156 & 37.86 & 11.3 & 9.86 & 7.19 & 10.59 & 7.99 &8.94 &  0.12843   \\
  157 & 39.03 & 10.54 & 9.54 & 6.85 & 9.73 & 8.89 & 8.61 &  0.58512   \\
  158 & 38.03 & 10.12 & 9.03 & 6.99 & 9.41 & 9.27 & 9.28 &  0.21385   \\
  159 & 38.19 & 11.09 & 9.72 & 6.77 & 10.35 & 8.64 & 8.79 &  0.1897   \\
  160 & 38.9 & 10.92 & 9.58 & 7.3 & 10.13 & 8.92 & 8.71 &  0.55596   \\
  161 & 38.46 & 11.68 & 9.28 & 6.3 & 11.17 & 7.52 & 8.45 &  0.00612   \\
  162 &  ---  & 11.46 & 9.22 & 6.37 & 10.66 & 7.29 & 8.34 &  0.00148   \\
  163 & 38.04 & 11.61 & 9.72 & 7.2 & 10.88 & 8.74 & 8.76 &  0.24571   \\
  164 &  ---  & 10.83 & 8.64 & 7.13 & 10.22 & 7.21 & 8.35 &  0.00056   \\
  165 & 38.57 & 9.98 & 8.84 & 6.53 & 9.32 & 8.41 &  ---  &  0.09959   \\
  166 &  ---  & 11.74 & 8.63 & 6.56 & 11.21 & 7.74 & 8.04 &  0.00265   \\
  167 & 38.15 & 10.72 & 9.04 & 6.82 & 10.12 & 8.00 &  ---  &  0.03628   \\
  168 & 38.82 & 9.89 & 8.88 & 6.34 & 9.2 & 9.29 &  ---  &   0.18107   \\
  169 & 38.46 & 10.33 & 8.87 & 6.43 & 9.26 & 8.48 &  ---  &   0.13606   \\
  170 & 37.96 & 11.62 & 9.55 & 7.38 & 10.93 & 8.65 & 8.78 &  0.07806   \\
  171 & 38.8 & 10.67 & 9.55 & 6.78 & 9.92 & 8.55 & 8.68 &  0.3173   \\
  172 & 38.01 & 10.69 & 9.21 & 6.71 & 9.94 & 7.91 & 8.67 &  0.00681   \\
  173 & 38.78 & 11.3 & 9.66 & 6.85 & 10.67 & 8.43 & 9.01 &  0.18459   \\
  174 & 38.57 & 11.45 & 9.21 & 6.25 & 10.95 & 7.82 & 8.62 &  0.00191   \\
  175 &  ---  & 11.17 & 8.32 & 6.67 & 10.59 & 7.22 & 7.82 &  0.00081   \\
  176 & 38.15 & 11.45 & 9.37 & 6.84 & 10.96 & 7.64 &  ---  &  0.00446   \\
  177 & 38.44 & 10.37 & 9.29 & 6.61 & 9.59 & 8.70 & 8.52 &  0.39342   \\
  178 & 37.47 & 12.25 & 9.16 & 6.96 & 11.72 & 7.81 & 7.63 &  0.02393   \\
  179 & 37.14 & 11.5 & 8.43 & 6.36 & 10.84 & 7.90 & 7.82 &  0.00126   \\
  180 &  ---  & 11.42 & 8.41 & 5.77 & 10.99 & 7.04 & 7.69 &  0.00085   \\
  181 &  ---  & 11.07 & 8.22 & 5.97 & 10.36 & 7.74 & 7.54 &  0.00087   \\
  182 & 38.54 & 10.48 & 9.32 & 6.84 & 9.71 & 8.89 & 8.35 &  0.31738   \\
  183 & 38.0 & 11.96 & 8.94 & 7.1 & 11.3 & 7.82 & 8.17 &  0.00245   \\
  184 &  ---  & 10.39 & 9.08 & 6.07 & 9.69 & 7.31 & 8.37 &  0.045   \\
  185 & 38.36 & 10.71 & 8.73 & 6.71 & 10.07 & 7.62 & 8.33 &  0.01417   \\
  186 &  ---  & 10.48 & 8.49 & 5.1 & 10.83 & 7.78 & 8.22 &  0.00347   \\
  187 & 38.33 & 10.64 & 9.7 & 7.3 & 9.79 & 9.47 & 8.61 &  1.06061   \\
  188 & 38.37 & 10.57 & 9.33 & 6.99 & 9.75 & 9.12 & 8.39 &  0.37532   \\
  189 & 38.68 & 10.07 & 8.99 & 6.53 & 9.34 & 8.39 &  ---  & 0.17429  \\
  190 & 39.35 & 11.94 & 10.51 & 8.09 & 11.24 & 9.30 & 9.50 &  2.52532   \\
  191 & 38.21 & 9.71 & 8.44 & 5.88 & 8.94 & 8.20 &  ---  &  0.0627   \\
  192 & 37.68 & 10.24 & 8.35 & 5.73 & 9.6 & 7.21 &  ---  &  0.00031   \\
  193 & 38.44 & 10.15 & 9.23 & 6.67 & 9.36 & 8.44 & 8.16 &  0.33063   \\
  194 & 38.41 & 11.45 & 10.17 & 8.11 & 10.7 & 9.97 & 9.13 &  1.22591   \\
  195 &  ---  & 10.47 & 7.85 & 5.83 & 9.78 & 7.97 & 8.44 &  0.00034   \\
  196 & 38.68 & 10.49 & 9.64 & 6.99 & 9.72 & 9.50 & 8.37 &  0.91001   \\
  197 & 38.6 & 10.5 & 9.32 & 6.9 & 9.73 & 8.69 & 9.02 & 0.27937  \\
  198 & 38.21 & 10.4 & 8.85 & 6.89 & 9.62 & 8.82 & 8.26 &  0.11905  \\
  199 & 38.5 & 10.13 & 8.77 & 6.25 & 9.38 & 7.91 &  ---  &   0.0751  \\
  200 & 39.16 & 11.81 & 9.8 & 7.01 & 11.14 & 7.44 & 8.99 &  0.00523  \\
  201 & 39.51 & 11.62 & 10.45 & 7.96 & 11.01 & 9.87 & 9.41 &  2.17945  \\
  202 &  ---  & 9.86 & 8.56 & 6.32 & 9.28 & 7.50 &  ---  &  0.03914  \\
  203 & 39.19 & 10.6 & 9.85 & 7.0 & 9.8 & 9.45 & 8.40 &  1.40335  \\
  204 & 38.81 & 11.49 & 10.29 & 7.97 & 10.81 & 9.69 & 9.33 &  2.93965  \\
  205 & 38.89 & 11.37 & 10.44 & 7.66 & 10.64 & 9.71 & 9.25 &  3.62826  \\
  206 & 38.23 & 10.14 & 9.1 & 6.41 & 9.56 & 7.98 & 8.96 &  0.15441  \\
  207 & 38.87 & 10.68 & 9.38 & 6.92 & 9.94 & 8.50 & 8.69 &  0.17767  \\
  208 & 38.29 & 11.54 & 9.79 & 7.71 & 10.9 & 8.84 & 8.61 &  0.41251  \\
  209 &  ---  & 11.34 & 8.17 & 6.46 & 10.57 & 8.25 & 7.49 &  0.00026  \\
  210 &  ---  & 10.92 & 7.91 & 5.36 & 10.28 & 7.97 & 7.82 & 0.00024  \\
  211 & 37.28 & 11.69 & 8.65 & 7.36 & 11.21 & 7.72 & 8.08 &  0.01085  \\
  212 & 38.64 & 10.27 & 9.05 & 6.43 & 9.35 & 9.4 & 8.80 &  0.26751  \\
  213 & 38.51 & 12.22 & 10.43 & 8.38 & 11.29 & 10.28 & 9.23 &  1.42139  \\
  214 &  ---  & 11.24 & 8.12 & 6.08 & 10.69 & 7.95 & 7.86 &  0.00163  \\
  215 & 39.33 & 11.03 & 9.85 & 7.22 & 10.2 & 9.03 & 9.67 &  0.57985  \\
  216 & 39.0 & 11.38 & 10.28 & 7.68 & 10.65 & 9.23 & 9.23 &  1.66949  \\
  217 & 37.81 & 11.73 & 10.17 & 7.65 & 10.96 & 8.88 & 9.35 &  0.02413  \\
  218 &  ---  & 11.29 & 8.32 & 6.48 & 10.72 & 7.43 & 7.91 &  0.00107  \\
  219 &  ---  & 11.05 & 8.54 & 6.64 & 10.38 & 7.22 & 7.85 &  0.00071  \\
  220 & 38.77 & 11.81 & 10.04 & 7.74 & 11.15 & 8.82 & 9.01 &  0.39101  \\
  221 & 38.5 & 10.81 & 9.34 & 6.95 & 10.09 & 7.60 & 8.59 &  0.0843  \\
  222 & 37.7 & 10.11 & 8.54 & 6.03 & 9.47 & 7.44 &  ---  &   0.00269  \\
  223 & 38.27 & 9.61 & 8.22 & 6.41 & 8.96 & 8.32 &  ---  &   0.0277  \\
  224 & 37.73 & 10.96 & 8.9 & 6.82 & 10.29 & 8.03 & 8.39 &  0.00116  \\
  225 & 37.82 & 9.82 & 7.79 & 5.75 & 9.04 & 7.83 &  ---  &   0.00799  \\
  226 & 38.34 & 10.29 & 8.94 & 6.48 & 9.59 & 8.38 &  ---  &   0.10411  \\
  227 & 38.12 & 10.34 & 9.23 & 6.98 & 9.56 & 9.95 &  ---  &   0.41618  \\
  228 &  ---     &  9.59 &  8.22  & 5.97 &  --- &   ---    &  ---  &  0.02221    \\ 
  229 &  ---  & 9.94 & 7.8 & 5.98 & 9.2 & 7.25 & 7.61 &  0.01798  \\
  230 & 38.48 & 10.33 & 9.17 & 6.68 & 9.59 & 8.68 & 8.55 &  0.23401  \\
  231 &  ---  & 11.44 & 8.44 & 5.68 & 11.01 & 7.11 & 7.94 &  0.00082  \\
  232 & 36.81 & 10.62 & 9.06 & 6.39 & 9.95 & 7.73 &  ---  &  0.00108  \\
  233 & 38.81 & 10.6 & 9.45 & 7.02 & 9.91 & 8.34 & 8.90 &  0.17121  \\
  234 &  ---  & 11.22 & 8.35 & 6.63 & 10.73 & 7.29 & 7.83 &  0.01427  \\
  235 &  ---  & 10.94 & 8.37 & 6.99 & 10.25 & 7.29 & 7.83 &  0.0006  \\
  236 &  ---  & 11.7 & 8.77 & 6.79 & 11.26 & 7.64 & 8.23 &  0.00799  \\
  237 & 38.79 & 10.44 & 9.47 & 6.73 & 9.7 & 8.71 & 8.82 &  0.43022  \\
  238 &  ---  & 9.71 & 8.3 & 5.8 & 8.84 & 9.24 &  ---  &  0.08862  \\
  239 & 38.77 & 10.7 & 9.73 & 7.01 & 9.92 & 8.71 & 8.90 &  0.56742  \\
  240 &  ---  & 11.09 & 8.02 & 5.88 & 10.4 & 7.41 & 7.76 &  0.00057  \\
  241 & 37.42 & 11.79 & 8.51 & 5.28 & 10.68 & 9.00 & 7.55 &  0.00163   \\
  242 & 39.31 & 10.89 & 9.45 & 6.92 & 10.18 & 9.11 & 8.49 &  0.4054   \\
  243 & 38.6 & 11.34 & 9.03 & 7.16 & 10.86 & 8.06 & 8.16 &  0.00324  \\
  244 & 39.3 & 10.97 & 9.92 & 7.33 & 10.26 & 8.73 & 9.25 &  0.81302  \\
  245 & 38.17 & 12.07 & 8.97 & 6.65 & 11.52 & 7.77 & 7.98 &  0.02284  \\
  246 & 38.96 & 11.19 & 9.92 & 7.43 & 10.47 & 9.63 & 8.83 & 1.11254  \\
  247 & 38.74 & 11.31 & 10.3 & 7.75 & 10.52 & 9.54 & 9.43 &  2.67483  \\
  248 &  ---  & 11.1 & 8.57 & 6.9 & 10.35 & 7.83 & 7.72 &  0.00066  \\
  249 &  ---  & 9.83 & 7.91 & 5.89 & 9.19 & 7.83 &  ---  &   0.00083  \\
  250 &  ---  & 11.43 & 8.36 & 6.47 & 10.76 & 8.33 &  ---  &  0.00076  \\
  251 & 39.47 & 11.88 & 10.68 & 8.11 & 11.09 & 9.95 & 9.56 &  4.95298   \\
  252 &  ---  & 10.36 & 9.14 & 6.66 & 9.55 & 9.18 &  ---  &  0.34431   \\
  253 &  ---  & 11.24 & 9.21 & 5.83 & 10.51 & 8.22 & 7.93 &  0.0712   \\
  254 & 38.9 & 11.0 & 9.76 & 7.47 & 10.5 & 8.76 & 9.22 &  0.86308   \\
  255 & 38.24 & 10.34 & 9.1 & 6.87 & 9.48 & 9.33 & 7.63 &  0.41334   \\
  256 &  ---  & 10.98 & 9.93 & 6.84 & 10.14 & 8.46 & 8.83 &  1.31365   \\
  257 & 38.41 & 11.37 & 9.25 & 7.44 & 10.69 & 9.31 & 8.43 & 0.00207   \\
  258 & 37.06 & 11.94  & 8.69 & 5.48 & 11.10 &  ---  &  8.15  &   0.000641          \\
  259 & 38.95 & 10.53 & 9.47 & 6.99 & 9.65 & 9.58 & 8.65 &  0.70417   \\
  260 & 39.23 & 11.38 & 9.78 & 7.03 & 10.76 & 7.76 & 8.51 &  0.00252   \\
  261 &  ---  & 10.34 & 8.78 & 6.87 & 9.59 & 8.73 &  ---  &   0.03521   \\
  262 & 38.72 & 10.65 & 9.74 & 7.07 & 9.8 & 9.51 & 8.80 &  0.9488   \\
  263 & 38.73 & 12.01 & 10.1 & 8.17 & 11.26 & 9.91 & 9.29 &  0.89124   \\
  264 &  ---  & 10.07 & 8.48 & 6.61 & 9.32 & 8.4 &  ---  &   0.02631   \\
  265 &  ---  & 10.24 & 9.11 & 6.30 & 9.05 &  ---  & ---  &   0.25801   \\
  266 & 37.95 & 10.78 & 9.74 & 7.56 & 9.9 & 10.15 & 8.69 &  1.37478  \\
  267 & 38.19 & 10.56 & 9.4 & 6.88 & 9.87 & 9.28 &  ---  &   0.30486  \\
  268 & 39.11 & 10.6 & 9.75 & 7.01 & 9.77 & 9.08 & 8.65 & 0.71238  \\
  269 &  ---  & 11.43 & 8.79 & 7.22 & 10.85 & 7.50 &  ---  &  0.00146  \\
  270 &  ---  & 11.79 & 9.58 & 7.06 & 11.02 & 8.09 & 8.24 &  0.00275  \\
  271 & 38.82 & 10.34 & 9.16 & 6.93 & 9.61 & 8.95 &  ---  &   0.12211  \\
  272 &  ---  & 11.49 & 8.48 & 6.45 & 10.81 & 7.99 & 8.03 &  0.00154  \\
  273 & 39.21 & 10.81 & 9.33 & 7.22 & 10.08 & 8.93 & 8.65 & 0.19931  \\
  274 & 38.71 & 11.06 & 8.99 & 6.92 & 10.37 & 9.01 & 8.45 &  0.01968   \\
  275 & 38.79 & 10.84 & 9.8 & 7.17 & 9.89 & 9.64 & 8.13 &  1.51157   \\
  276 & 38.6 & 10.51 & 9.36 & 6.65 & 9.69 & 8.78 & 8.94 & 0.35483   \\
  277 &  ---  & 9.97 & 8.87 & 6.26 & 9.84 & 7.77 &  ---  &   0.08273   \\
  278 & 37.63 & 10.25 & 8.59 & 6.25 & 9.59 & 7.65 &  ---  &   0.01246   \\
  279 & 37.93 & 10.52 & 9.07 & 7.1 & 9.66 & 9.21 &  ---  &   0.26801   \\
  280 & 39.05 & 10.48 & 9.2 & 6.53 & 9.76 & 8.62 & 8.53 & 0.13179   \\
  281 & 38.71 & 9.64 & 9.06 & 6.65 & 9.62 & 8.13 &  ---  &  0.17645  \\
  282 &  ---  & 10.15 & 7.99 & 6.18 & 10.29 & 7.85 & 7.71 &  0.00136   \\
  283 & 39.05 & 10.77 & 9.89 & 7.21 & 9.95 & 9.68 & 8.61 &  1.1173  \\
  284 &  ---  & 10.85 & 9.65 & 7.16 & 10.2 & 8.93 & 8.89 &  0.25598  \\
  285 & 38.23 & 11.32 & 9.95 & 7.34 & 10.66 & 8.24 & 8.77 & 0.22755  \\
  286 &  ---  & 11.22 & 8.61 & 6.88 & 10.68 & 9.30 & 8.48 &  0.00371  \\
  287 & 39.08 & 10.64 & 9.54 & 6.96 & 9.86 & 8.99 & 8.61 &  0.48475  \\
  288 & 38.43 & 11.01 & 9.35 & 7.14 & 10.29 & 8.51 & 8.40 &  0.18297  \\
  289 & 39.49 & 11.33 & 10.02 & 7.45 & 10.46 & 9.68 & 8.93 &  1.35923  \\
  290 & 39.33 & 10.34 & 9.2 & 6.17 & 9.48 & 8.58 & 8.66 & 0.16195  \\
  291 &  ---  & 10.68 & 7.68 & 5.92 & 9.98 & 8.43 & 7.62 &  0.00057  \\
  292 & 39.41 & 10.68 & 9.62 & 6.9 & 9.95 & 9.02 & 8.96 &  0.42406  \\
  293 & 39.01 & 10.4 & 9.38 & 6.81 & 9.53 & 8.96 & 8.25 & 0.47436  \\
  294 & 38.44 & 10.52 & 9.33 & 6.91 & 9.75 & 8.89 &  ---  &  0.18826  \\
  295 & 39.63 & 11.23 & 10.37 & 7.78 & 10.67 & 9.67 & 9.61 &  2.79919  \\
  296 & 37.96 & 10.88 & 8.48 & 5.89 & 10.16 & 7.82 & 7.62 &  0.00045  \\
  297 & 38.75 & 11.04 & 9.77 & 7.51 & 10.27 & 9.55 & 8.75 &  0.69257  \\
  298 & 39.69 & 10.45 & 9.58 & 6.6 & 9.55 & 8.98 &  ---  &  0.62548  \\
  299 & 38.63 & 10.92 & 9.25 & 7.24 & 10.0 & 9.04 & 8.25 &  0.3715  \\
  300 & 38.2 & 10.46 & 8.6 & 6.16 & 9.68 & 8.39 &  ---  &   0.01081  \\
  301 & 38.65 & 11.02 & 9.37 & 7.49 & 10.09 & 9.48 & 8.47 &  0.45455  \\
  302 & 38.38 & 10.28 & 8.9 & 6.9 & 9.5 & 9.24 &  ---  &   0.11592  \\
  303 & 39.82  & 10.40 & 9.60 & 6.46 & 9.23 &  ---   &  ---   &   0.568607   \\
  304 & 38.02 & 10.66 & 9.25 & 7.16 & 9.95 & 8.64 & 8.62 &  0.08512  \\
  305 & 38.13 & 9.99 & 8.16 & 5.97 & 9.28 & 7.60 &  ---  &   0.01451  \\
  306 &  ---  & 11.6 & 9.12 & 6.63 & 11.04 & 8.06 & 8.57 &  0.00233  \\
  307 & 38.74 & 11.42 & 9.86 & 7.84 & 10.69 & 9.59 & 8.44 &  1.08746  \\
  308 & 39.46 & 9.43 & 7.61 & 5.22 & 8.71 & 7.89 &  ---  &  0.00281  \\
  309 & 38.29 & 9.99 & 8.81 & 6.3 & 9.11 & 9.08 &  ---  &   0.16234  \\
  310 & 38.57 & 10.74 & 9.63 & 7.01 & 9.99 & 9.01 & 8.84 &  0.42236 \\
  311 & 38.76 & 11.64 & 9.69 & 7.61 & 10.92 & 9.08 & 8.81 & 0.13983  \\
  312 &  ---  & 11.46 & 8.35 & 6.1 & 10.74 & 7.45 & 7.99 & 0.00087  \\
  313 & 38.4 & 10.69 & 9.26 & 7.22 & 9.89 & 9.05 & 8.32 &  0.20532  \\
  314 & 38.6 & 10.39 & 9.18 & 6.82 & 9.49 & 9.27 &  ---  &  0.38941  \\
  315 & 37.36 & 9.89 & 8.56 & 6.37 & 9.08 & 9.30 &  ---  &   0.13785  \\
  316 &  ---  & 11.39 & 8.29 & 6.4 & 10.66 & 7.96 & 7.84 &  0.00145  \\
  317 &  ---  & 9.97 & 8.46 & 6.41 & 9.2 & 8.82 &  ---  &  0.05836  \\
  318 & 38.68 & 10.67 & 9.42 & 6.91 & 9.77 & 9.29 & 8.52 &  0.47551  \\
  319 & 38.52 & 10.67 & 9.5 & 7.26 & 9.86 & 9.53 & 8.37 &  0.80481  \\
  320 & 38.72 & 10.8 & 9.73 & 7.52 & 10.0 & 9.68 & 8.35 &  1.25507  \\
  321 &  ---  & 10.41 & 9.34 & 6.5 & 9.55 & 8.66 &  ---  &  0.36288  \\
  322 & 38.43 & 11.41 & 9.08 & 7.4 & 10.66 & 9.85 &  ---  &  0.16154  \\
  323 & 39.16 & 10.89 & 9.64 & 7.16 & 10.09 & 9.36 & 8.72 & 0.32984  \\
\end{longtable}
\label{tabledata}
\twocolumn

\section{Mathematical tools}\label{tools}

\subsection{Estimation of a bivariate luminosity and mass function using a semi-parametric approach}\label{blf}

To estimate the bivariate Probability Distribution Function (PDF), $\psi(x,y)$, \citet{PaperI} have used a procedure based on the {\rm copulas} \citep[see][for the mathematical definition]{sch07}.
The PDF is derived from a given a set of $N$ {\it observed} quantities $\{ x_i \}_{i=1}^N$ and $\{ y_i \}_{i=1}^N$ such that $\psi(x,y) dx dy$
is the probability that a random variable (in this case the luminosity or the mass) takes values in the range $[x, x+dx]$ and $[y, y+dy]$.

The method requires the computation of the cumulative distribution function (CDF) of the PDFs $\phi(x)$ and $\theta(y)$ (hereafter called {\it marginals}), defined from the following equations:
\begin{align}
\Phi(x) & = \int_{x'_{{\rm min}}}^x \phi(x') dx', \label{eq:Phi} \\
\Theta(y) & = \int_{y'_{{\rm min}}}^y \theta(y') dy', \label{eq:Theta}
\end{align}
which are distributed according to a uniform distribution that takes values in the range $[0, 1]$.
Defining $u_x=\Phi(x)$ and $u_y=\Theta(y)$, and if $G^{-1}(u_z)$ is the inverse function of the standard Gaussian CDF
$G(z)$, the quantities $z_x$ and $z_y$:
\begin{align}
z_x & = G^{-1}(u_x), \label{eq:z1} \\
z_y & = G^{-1}(u_y), \label{eq:z2}
\end{align}
are distributed according to a standard Gaussian PDF, $g(z)$; i.e., they are Gaussian variables. In other words, by means of Eqs.~(\ref{eq:Phi})-(\ref{eq:z2}) the random variables $x$ and $y$ are  {\it Gaussianised}.
It is assumed that  the joint PDF $g_{\Sigmab}(z_x,z_y)$ of 
$z_x$ and $z_y$ is the bivariate Gaussian PDF with covariance matrix $\Sigmab$ given by
\begin{equation} \label{eq:Sigma}
\Sigmab=\left(
\begin{array}{cc}
1 & \rho \\
 \rho & 1
\end{array}
\right),
\end{equation}
where $\rho$ is the linear correlation coefficient of the two random variables $z_x$ and $z_y$
(see \citet{tak10}).

The {\rm copula}  $C_{\Sigmab}(u_x, u_y )$ of $g_{\Sigmab}(z_x,z_y)$ is defined from the equation \citep[i.e.][]{sch07}:

\begin{equation} 
\psi \left(x, y\right) = c(u_x, u_y) \phi\left(x\right) \theta\left(y\right), \label{eq:pdf}
\end{equation} 
where
$x=\Phi^{-1}(u_x)$ and $y=\Theta^{-1}(u_y)$ and 
\begin{equation} \label{eq:c}
c_{\Sigmab}(u_x, u_y) = \frac{\partial^2 C_{\Sigmab}(u_x, u_y)}{\partial u_x \partial u_y}.
\end{equation}

We recall that a $d$-dimensional {\rm copula} $C: [0, 1]^d \rightarrow [0, 1]$ is a CDF with uniform marginals. 
{\rm Copulas} are used to describe the dependence between random variables, and their main use is to disentangle marginals and the dependence structure.
In particular, with the Gaussian {\rm copula} the dependence structure is parametrised by a single parameter, the correlation coefficient.

It is possible to see that
\begin{equation} \label{eq:C}
C_{\Sigmab}(u_x, u_y ) = G_{\Sigmab} \left( g^{-1}(u_x), g^{-1}(u_y) \right),
\end{equation}
with $ G_{\Sigmab}$ the CDF of the  bivariate Gaussian with covariance matrix $\Sigmab$,
from Eq.~\eqref{eq:c} it is
\begin{equation} \label{eq:cg}
c_{\Sigmab} (u_x, u_y) = \frac{1}{| \Sigmab|} \exp{ \left\{ - \frac{1}{2} \left[ \Gb^{-T} (\Sigmab^{-1} - \Ib) \Gb^{-1}\right] \right\}}. 
\end{equation}
Here, $\Gb^{-1} \equiv \left( G^{-1}(u_x), G^{-1}(u_y) \right)^T$, $\Gb^{-T}$ is the transpose of $\Gb^{-1}$, $\Ib$ the identity matrix and  $|\Sigmab|$ the determinant of $\Sigmab$.
In summary, to obtain a full description of the two variables together two ingredients are needed: the marginals and the type of interrelation.

Using the above results, a procedure for estimating the bivariate PDF $\psi(x,y)$ in the presence of possible left-censored data (upper limits) is the following.
\begin{enumerate}
\item Estimation of the marginals $\phih(x)$ and $\thetah(y)$ 
\item Computation of the uniform random variates/upper limits  $u_{x_i} = \Phih(x_i)$, $u_{x_j} = \Phih(x_j)$,  $u_{y_k} = \Thetah(y_k), $ and  $u_{y_l} = \Thetah(y_l)$ by means of Eq.~(\ref{eq:Phi})-(\ref{eq:Theta}); \\
\item Computation of the standard Gaussian variates/upper limits  $z_{x_i}$, $z_{x_j}$, $z_{y_k}$ and $z_{y_l}$ by means of Eqs.~(\ref{eq:z1})-(\ref{eq:z2});\\
\item Maximum Likelihood estimation of the linear correlation coefficient and then of matrix $\Sigmab$; \\
\item Computation of $\psi(x,y)$ for specific values of $x$ and $y$ by means of Eqs.~(\ref{eq:pdf})-(\ref{eq:cg}).
\end{enumerate}
The {\it copula} related to $z_{x_i}$, $z_{x_j}$, $z_{y_l}$, and $z_{y_k}$ is the same as the one related to $x_i$, $x_j$, $y_k$, and $y_l$. This is due to the {\it invariance property} of {\it copulas} by which the dependence captured by a {\it copula} is invariant with respect to increasing and continuous transformations of the marginal distributions \citep[see page 13 in][]{tri05}.

The procedure of this method, with its semi-parametric solution, as outlined in detail above it is extensively discussed in \citet{PaperI}.

\section{The bivariate luminosity and mass functions}\label{results}
The first step in the above procedure is the estimation of the marginals. In absence of any a priori knowledge of the analytical form of  $\phih(x)$ and $\thetah(y)$, a possible approach is represented by the system of three families,
say SU, SB, SL,  introduced by \citet{joh49} according to the fact that a random variable is unbounded (SU), bounded above and below (SB) or bounded only below (SL).
A detailed description of such families as well of their use in practical  application is given in \citet{vio94}. Here, it is sufficient to say that the members of this system are characterised by four free parameters
that allow them a great flexibility in reproducing most of the classical PDFs. As described in \citet{vio94}, a robust method to select the specific family and to estimate the corresponding parameters for a given set of data is based 
on the percentiles of their empirical distribution. In the present case, this method indicates the SB family 
\begin{equation}
f(x) = \frac{\eta}{\sqrt{2 \pi}} \frac{\lambda}{(x-\epsilon)(\lambda-x+\epsilon)} \exp{-\left\{\frac{1}{2}\left[ \gamma + \eta \ln\left(\frac{x-\epsilon}{\lambda - x + \epsilon}\right) \right]^2\right\}}
\label{SB}
\end{equation}
with $\epsilon \le x \le \lambda + \epsilon$
as the most suited to reproduce the PDF of the observed data. Table~\ref{tab:coeff} shows the estimated parameters and Fig.~\eqref{fig:histograms} 
the corresponding PDFs vs the experimental histograms.  Finally,
Figs~\eqref{SMF}-\eqref{MH2MF} show the corresponding bivariate PDF obtained by means of Eq.~\eqref{eq:pdf}.

It is important to stress here that, as explained in detail in \citet{PaperI}, we have considered several PDFs to fit the \Kb~ luminosities. All of them have a support of type 
$L_{\rm min} < L < \infty$ (or $M_{\rm min} < M < \infty$), a steep slope for $L \to L_{\rm min}$ (or $M \to M_{\rm min}$) and the possibility that $\phi(L), \phi(M) \to \infty$.  But since the $L_{\rm min}$ (or $M_{\rm min}$) is unknown, the three-parameters version of such PDF has to be used.
The fit of this kind of PDFs is a difficult problem since the maximum likelihood approach fails if $\phi(L), \phi(M) \to \infty$ when $L \to L_{\rm min}$ (or $M \to M_{\rm min}$). 

This problem has been solved with  the method described in Appendix~A  in \citet{PaperI}.

\begin{table*}
\noindent \centering{}\protect\caption{Coefficients of the PDFs of the SB family (eq.~\ref{SB}) for LTGs only and the whole sample \label{tab:coeff}}
\begin{tabular}{lccccc}
\hline \hline

Variable  &  $\eta$ &  $\epsilon$ & $\lambda$ & $\gamma$ 
 & corr \tabularnewline
\hline 
$L({\rm K}$)  & 0.6641 &	 	  0.002039    &  2.658   &    2.521     & 1.000 \tabularnewline
$L({\rm IR}$)  & 	0.7509 	 & 	 -0.001622  &     8.233    &   2.705 &      0.790 \tabularnewline
$M_{\rm dust}$  &   0.7874   & 	 -0.000153     &   6.773   &     3.482   &     0.849 \tabularnewline
$M_{\rm star}$  & 	0.6974    & 	  0.003965    &    2.879  &      2.467  &      0.967 \tabularnewline
SFR &       0.6316       &    -0.001695 &     11.15   &   2.498   &   0.389  \tabularnewline
$M_{\rm HI}$  & 0.6393   	 &   0.000966    &    2.519    &    2.277   &     0.431    \tabularnewline
$M_{\rm H2}$  & 0.9067   	 &   0.007500    &    14.85   &     2.881   &     0.672 \tabularnewline
$M_{\rm star} ^b$ 	& 0.6270         &   0.004132   &   8.343   &   2.682  &    0.965 
\\
\hline 
\\
\noindent
$^b$ computed over the whole sample\\

\end{tabular}
\end{table*}

\subsection{Analysis of the bivariate LFs}\label{BLFs}

Before discussing the outcomes of our analysis of the bivariate PDFs shown in Figs~\eqref{SMF}-\eqref{MH2MF} we examine the distribution
of the physical quantities listed in Table~\ref{tabledata}  and plotted in Figure~\ref{correlations}.

The green colours correspond to the galaxies classified as LT according to the classification in \citet{cor+12a}, while blue to the ET\footnote{The original classification was a NED-based morphological type classification \citep{bos+10} which has been modifed for several galaxies after revision based on more recent literature and visual ispection
as discussed in \citet{cor+12a}}.\hfill\break

The same colour code is used in the diagonal inlayings which contain the histograms of the two populations.
The first result is that the histograms are quite distinct and clearly show in most of the cases
the existence of two distinct distributions for LTGs and ETGs. The two populations cover different values of luminosities and masses.
Atomic, molecular and total gas masses, infrared luminosities and SFR are much larger in LTGs, while the \Kb~ luminosity and the stellar masses are larger in the ETGs.
The distribution of the 
dust mass is similar in LTGs and ETGs, but the distribution is shifted to lower value in ETGs of 0.5 dex.
The relations among these variables are not affected by the luminosity-distance relation,
because the HRS is a volume-limited sample. The values of the correlation coefficients are listed in the sixth column of Table~\ref{tab:coeff}.

The analysis of the correlations implies that a correct statistical analysis for many physical quantities can only be carried over a subset of the sample: either containing only ET or only LT. 
We have then computed the bivariate functions over the whole sample only in one case (stellar mass) 
for which we cannot distinguish the behaviour of the two populations.
In the following we compute most of the bivariates for LTGs, and, because of the limited statistics, only the overall trends are discussed for the ETGs in the sample.

Errorbars are computed with a bootstrap technique, iteratively extracting the values of the marginal functions (see \S~4 above) when the variable changes within its errorbar.

\subsubsection{The bivariate $L(K)-M_{\rm star}$}\label{LK-M*} 

Both Figure~\ref{correlations} and the bivariate LF ${L_K}$-${M_{\rm star}}$ in Figure~\ref{SMF} highlight
the tight relation between the \Kb~ luminosity and the stellar mass \citep{gav+96}. The correlation coefficient is 0.965. This is not surprising as 
the stellar masses are derived from the $i$-band luminosity and $g-i$ colour, close to the \Kb~ \citep{cor+12a, bos+09}.
Both ETGs and LTGs follow the same correlation, with the former objects containing larger stellar masses.
A few sources can be identified both in Figure~\ref{correlations} and Figure~\ref{SMF} as outliers because they deviate from the tight correlation.
This deviation is clearly visible in Figure~\ref{SMF} where the values are on a linear scale.

The \Kb ~~magnitude of the HRS sample are taken from the 2MASS survey, these outliers may have not well derived values of the  \Kb ~~mag due to their complex morphology and low surface brightness \citep{koc+01}, and closeness to the completion limit of the survey \citep{2MASS}.
Indeed, the errorbars on the photometry for the extended (and more nearby) and low surface brightness galaxies are more affected by sky fluctuations on large scale \citep[see discussion in Appendix A in][]{koc+01}.

Overall this correlation confirms a well known result that the  \Kb ~~mag is a fair tracer of the stellar mass.
Figure~\ref{SMF} shows the marginal function used to compute the bivariate in section~\ref{tools} representing the stellar mass distribution. This can be retrieved from the analytical form of equation~(\ref{SB})
using the parameters in Table~\ref{tab:coeff}. 
Bearing in mind that a strict comparison with previous works on the stellar mass function (SMF) of the local universe is not straightforward because our different way of computing this function, we draw in Figure~\ref{SMF} together with our derived marginal the Schechter \citep{sche76} functions whose parameters are extracted from the best fit of the SMFs by \citet{bal+12}. 
These authors have characterised the SMF at $z\sim0$ down to $M=10^8  M_\odot$ using data from the 
GAMA surveys and fit their data with a double Schechter \citep{sche76} function with a single value for the break mass, and provide a good fit to the data for $M>10^8  M_\odot$. They claim that this is approximately the sum of a single Schechter \citep{sche76}
function for the blue population and double Schechter function for the red population.
 
The deviation of the marginal derived in this work with the SMFs computed by \citet{bal+12}
at stellar mass lower than $\sim 3\times 10^9 {\rm M}_\odot$ simply highlights the incompleteness of the
HRS sample at low stellar mass, i.e. the lack of galaxies with low stellar mass in the HRS sample. The HRS sample covers a small volume and lacks the necessary depth to detect 
 \Kb ~~mag faint and/or low surface brigthness galaxies.\hfill\break
The low end of the mass function is dominated by disc galaxies, i.e. the LTGs, \citep{tha+16}, which the HRS misses as discussed above.

The high-mass end of the SMF is in agreement among the different samples and confirms previous findings that it is dominated by spheroidal galaxies. \citet{tha+16} show that at masses lower than log($M/M_\odot$)=10.3 there is a preponderance of disc galaxies, whereas increasing galaxy stellar mass, this disc dominance gradually decreases
with a corresponding increase in the spheroidal contribution. Although the SMF of the disc components shows a steep increase at the faint end, their contribution to the total galaxy stellar mass density
is only 37 per cent. 
This gradual change of the SMF from disc dominated to spheoridal dominated galaxies is linked to the physics and
environmental processes which drive the build-up of stellar mass in these two principal galaxy components \citep{tha+16}. 
The lack of small disc galaxies could be also due to the environment and this could be well the case for the HRS sample \citep{bos+14a,bos+16}. 
Changes on the shape of the local SMF (and in the value of M$_\star$) in the highest density environments, which contain an enhancement of massive galaxies,
are discussed in \citet{bla+09}.

\subsubsection{The bivariate $L(K)-L(IR)$}\label{LK-LIR}

The relation between the IR luminosity and the \Kb~ luminosity clearly highlights a dichotomy for the two
morphological types (ET) and  (LT), the relation is almost linear in logarithmic scale ($ L(K)\sim \alpha \cdot L(IR) $, with $ \alpha \sim 10,1000$
for LT and ET  respectively).
The IR luminosity is fainter in ETGs, while LTGs have fainter values of the \Kb~ luminosity.
A closer look at the \Kb~/IR luminosity relation in Figure~\ref{correlations} shows that one tenth of the objects have properties in between the relations defined for ET and LT respectively.
The majority of these objects host a weak AGN and/or are classified as 'retired galaxy' by \citet{gav+18}. These galaxies have been star-forming in the past and, although the nucleus is sterilised,
there are still remnants of star formation in the outer region. They share most of the properties of the ETGs, with less gas and very low specific SFR.
The lower left part of the  $L(K)-L(IR)$ plane is occupied mosty by objects classified as LTGs dominated by HII regions
\citep{gav+18}.

Because of the low number of ET  objects and the distinct behaviour of the ET and LT galaxies with respect to the IR Luminosity,
Figure~\ref{LIRLF} shows the BLF for LTGs only. In spite of the relative good correlation measured (0.79) the BLF shows a spread at large values of the IR and \Kb~ luminosities. The objects responsible for this spread
are galaxies of the Virgo Cluster and host a weak AGN and/or are classified as 'retired galaxy' by \citet{gav+18}. These objects might be in migration from the blue cloud (star-forming) to the green valley (post-starburst) and eventually to the
red sequence. Particulary those in the Virgo cluster loose gas (mainly atomic) through ram pressure stripping and have more compact discs because of the further loss of dust and molecular gas \citep{bos+14c,bos+16}.

In \citet{PaperI} a similar BLF has been computed limited to the monochromatic luminosities in the far-IR (in the $SPIRE$ bands at 250, 350 and 500$\mu$m).
\citet{PaperI} discussed how for LTGs the dependence  $L(K)-L(IR)$
can be interpreted as a physical connection between the cold component of the dust -- closely related to the galaxy dust mass -- and the stellar mass -- inferred from the  \Kb~ absolute luminosity -- which is a tracer of the mass of the old stellar population.

\noindent
Figure~\ref{LIRLF} shows the analytical form (the marginal) computed in section~\ref{tools} of the local IR luminosity function.
To locate this finding in the context of other authors' results, Figure~\ref{LIRLF} reports the best fit of the modified Schechter function derived by \citet{mar+16} using the \HER~HerMES survey.
This latter is the most recent version of the local IRLF based on blind FIR/submm surveys \citep{vac+10,neg+13,cle+13,wan+16} and computed using the total far-IR (3-1000$\mu$m) luminosity combined with models.

The agreement shown in Figure~\ref{LIRLF} is good with small deviations at low and large luminosities. While at large luminosities the HRS is statistically not complete because of the small surveyed area, the
discrepancy at low IR luminosities can be attributed to the ways these samples have been selected: the HRS sample is a \Kb~ selected and may miss very low IR luminosity objects which are more easily detected in blind FIR surveys.
There might be an additional factor related to the definition of morphology/colour in the infrared and in the optical. \citet{mar+16} interpret the shape of the local far-IR LF as due to the contributions of red (possibly ET)
and blue (possibly LT) galaxy populations, with their different Schechter forms, rapidly evolving already at low redshifts.
However, the cut-off line between red and blue galaxies in this context is less sharp than in the optical classification of the galaxy morphology, as in the HRS sample, where among red galaxies there are red spiral galaxies that could be the result of their highly inclined orientation and/or a strong contribution of the old stellar population \citep[see also][]{dar+16}.

\subsubsection{The bivariate Dust Mass Function (DMF), $L(K)-M_{\rm dust}$, and the relation $M_{\rm dust}-M_{\rm star}$}\label{LK-Md}

As shown in Figure~\ref{correlations} the variation of the  \Kb~ luminosity in ETGs is roughly constant with respect to the dust mass, and, considering also the upper limits, there is no correlation between the star luminosity and the dust emission.
For this reason and for the low statistical significance of the number of ETGs we compute the bivariate PDF $L_K-M_{\rm dust}$ shown in Figure~\ref{MDF} only for LTGs.
The dependence of the \Kb~ luminosity and the dust mass shows a very tight relation, with a correlation coefficient of 0.849, slightly stronger than that between 
$L(K)-L(IR)$. This is expected as the dust thermal emission is the main contributor of the IR luminosities.

Stellar and dust masses seem to be in tight relationship, which could be interpreted as a relationship among the stellar mass, the cold dust mass, and the far-infrared luminosity in LTGs. This tightness can alse be due to the presence of old stars which dominate the stellar mass and at the same time producing the dust in their stellar winds
\citep{dwe+98,zhu+08,zhu+16}.
What is clear is that the distribution of the cold dust in the galaxy discs follows that of the stars.

The spread of this relation, due to the objects with larger dust mass and low \Kb~ luminosity (small stellar mass), can be ascribed to the fact that lower (stellar) mass galaxies have higher dust mass fractions than their more massive counterparts \citep{cor+12b,cle+13}.

From this function, using equation~(\ref{SB}) and Table~\ref{tab:coeff}, 
we can derive the DMF which is shown in Figure~\ref{MDF}. In the same figure we plot the DMFs obtained by \citet{dun+11}, the best-fist model by \citet{cle+13} 
and the recent one by \citet{bee+17}. \citet{dun+11} and \citet{bee+17} compute the DMF over a sample of \HER ~selected galaxies, while \citet{cle+13} over a sample selected from the $Planck$ source catalogue.
\citet{cle+13} claim agreement with the \citet{dun+11} values because of the large uncertainties on the derivations of the dust masses, which are mainly linked to the assumed physical properties of the dust
 \citep[see also][]{dev+17}. 
All these models make use of SED fitting templates to derive the physical parameters of the galaxies (therefore also the value of the dust mass).
The mostly used MAGPHYS package \citep{cle+13,bee+17,dri+18} combines black bodies with different temperatures, keeping the energy balance between UV/optical and NIR while for HRS \citet{cies+14} fitted the Draine and Li models only on the IR part.\hfill\break
Although the accuracy of the dust mass values mainly depends on the
quality of the fit (i.e. the number of photometric points), it is also largely depending on the dust model, which assumes dust absorption coefficient differing up to a factor of two, among the different models
Other uncertainties may arise from the selection criteria and systematics which are not perfectly under control.

To overcome some of the discrepancies we make use of the recomputed values reported in \citet{bee+17} who have rescaled the DMFs at the same value of the dust absorption coefficient. These rescaled DMFs are those shown in Figure~\ref{MDF}.

The DMF computed for the HRS sample lies in between the DMF given by \citet{cle+13}'s and those derived by \citet{dun+11,bee+17}.
We do not want to overinterpret this result because of the difference on the dust models, the difference in the selection wavelengths (250$\mu$m and 500$\mu$m), and in the catalogues ($Herschel$/SPIRE and $Planck$). We can claim that the Local DMF derived from the HRS is consistent with the values found for far-IR/submm galaxies. We need to keep in mind, however, that
the HRS may miss a number of dusty galaxies because it targets \Kb~ selected objects.

\HER ~galaxy samples contain red galaxies which may correspond to the optical classification of both LTGs and ETGs (i.e. contain part of the ETGs of the HRS sample), with dust masses similar to the blue objects, i.e. normal spiral/star-forming systems. Some red ETGs keep the properties of optical ETGs (lower mean dust-to-stellar mass ratios, lower mean star-formation / specific-star-formation rates) but
a population of ETGs exists, containing a significant level of cold dust similar to those observed
in blue/star-forming galaxies. The origin of dust in such ETGs it is still unclear. It could be of external origin (e.g. fuelled through mergers
and tidal interactions, \citet{dar+16}) or long lived in galaxy discs, with late results favouring this latter interpretation \citep{bas+17}, \citep[see also][]{gom+10,cor+12b,smi+12,agi+15,eal+18}.

The tight relation between the \Kb~ luminosity and the stellar mass (see Figure~\ref{correlations} and in Figure~\ref{SMF}) allows to explore as well the relation between the dust and the stellar masses. 
This latter is very tight for the LTGs, while no clear connection is detected in ET objects. This is expected as about half of the ETGs remain undetected in the \HER~bands \citep{cor+12b,smi+12}, the corresponding IR luminosity and derived dust masses have to be considered as upper limits \citep{cies+14}.

\citet{cor+12b} show that the spread in the relation between stellar and dust masses in the HRS may be attributed
to the variation of the dust content as a function of the environment and of the HI content more than to the morphological (late versus early) type.

\subsubsection{The bivariate $L(K)$-{SFR}}\label{LK-SFR}

The SFR BLF is displayed in Figure~\ref{SFRLF}. The computed bivariate function shows a slight relation between the SFR and the \Kb~ luminosity, with this latter, as highlighted in Figure~\ref{correlations} and in Section \S ~\ref{LK-M*} strongly linked to the stellar mass.

Figure~\ref{SFRLF} displays the SFR functions derived in this work from the BLF and compares it with the values obtained from other samples. The comparison is not straightforward because of the way the SFR has been computed in the different samples.
The HRS SFR is the average value among that derived from the dust corrected H$\alpha$ luminosity, the far-UV dust corrected luminosity and the radio emission at 20cm \citep{bos+15}.
SFR values in other samples have been obtained either from the
H$\alpha$ measurements alone \citep{bot+11,gun+13} or translating the IR luminosity to SFR \citep{cle+13}.

It is straightforward to see that the SFR function is a strong function of the sample selection criteria. While the SFR function extracted from the H$\alpha$ is significantly lower than that 
computed from the IR luminosities the behaviour of the HRS SFR function misses large values of the SFR. 

We are not at all astonished to see a large difference at high star formation rates 
with the Planck derived \citet{cle+13}'s LF, which includes FIR selected starbursts known to be totally 
absent in the local universe (within 25 Mpc the most extreme case is M82, with $\sim$10M$_\odot$/yr) and might be 
limited/biased by confusion. Furthermore, the difference in the several published $H_\alpha$ selected 
SFR LF is huge \citep[see Fig. 11][]{bos+15}, even within the same work once different samples are used. \citet{gun+13} published two different SFR LFs derived from $H_\alpha$, one from SDSS data, the second one from GAMA data, which is higher at least at low SFR values.
\citet{bos+16} have compared these  $H_\alpha$ LF (GAMA and SDSS) to the one derived using NUV data 
in the Virgo clusterv periphery and they match pretty well. This means that the observed differences in the SFR LF between HRS and \citet{gun+13}'s are mainly due to the sample, and not to the method to derive SFR.

In the past two decades a vast number of works have investigated the link between the SFR and the stellar mass \citep[i.e.][and references therein]{eal+17}.
Our interpretation of the SFR bivariate function is that the relation between
the SFR and $M_{\rm star}$ is a combination of at least two factors.
On the one hand, there is an effect due to the environment. \citet{bos+16} link the decrease in the star formation activity in the main sequence relation to HI-deficiency, which may be due to ram pressure stripping
\citep{bos+15,bos+16}. 
The location of the galaxy main sequence is different for objects which do show sign of perturbation from that drawn by unperturbed systems.
Many of the HRS galaxies show sign of perturbation due to the environment and a large infall rate of star forming systems is observed in Virgo.

On the other hand, there is a selection effect. The HRS sample contains most of the stellar mass in a specific volume of local Universe and, as discussed above in \S~\ref{LK-Md}, it should not be biased towards galaxies with high star formation rates. But it contains optically classified red galaxies that are red not only because of the old stellar population but because of a fraction of dust and gas
which show that they are still forming stars. 30\% of the red population classified as ET still contains a fraction of dust and have a residual star formation rate \citep{eal+17}. 
For very red objects, those with the lowest values of the SFR the redness is due to an old population and not to dust reddening and the values of the ratio $\frac{M_{\rm dust}}{M _{\rm star}}$ are $<10^{-4}$.

We are not able to investigate further this issue using the BLF. The number of objects is low to split the sample and compute the BLF differently on the galaxies belonging to the cluster and those of the fields.

\subsubsection{The bivariates $L(K)-M_{gas}$, atomic and molecular gas}\label{LK-Mgas}

In Figures~\ref{correlations},~\ref{MHIMF},~\ref{MH2MF},
we report the values of the distributions of the atomic gas, and the molecular gas masses and the bivariate mass functions. The amount of atomic and molecular gas is a strong function of the morphological type, where most of the ETGs are undetected in atomic and molecular gas \citep{bos+14a}.

The derived atomic gas function for LTGs only and those obtained by HI dedicated surveys \citep{zwa+05,mar+10,hop+15,jon+18} are shown in Figure~\ref{MHIMF}.
Overplotted are also the predictions by \citet{pop+14} and \citet{lag+11}.
The $M(HI)-MF$ derived from the HRS data differs substantially at values $M(HI)>$ a few $\times10^9$M$_\odot$.
The weak correlation of the HI mass with the \Kb~ luminosity  (see Figure~\ref{MHIMF}a and Table~\ref{tab:coeff})
does not allow to strongly constrain the bivariate and as a consequence the construction of the atomic gas mass function is poorly determined. This may explain the strong difference in shape observed in the HI MF.

The deficiency of large mass objects can be explained twofold: the HRS is a \Kb~ luminosity selected, while the HI dedicated surveys are blindly selecting HI emitting galaxies \citep{zwa+05,mar+10,hop+15,jon+18}.
The HRS sample therefore may miss most of the HI-massive galaxies. Secondly, the HRS contains more HI deficient objects as normal field galaxies (roughly half of the sample).
This fact is attributed to the presence of the Virgo cluster and its gravitational effect on the gas. Through direct stripping of the ISM from the disc (e.g., ram pressure) the galaxy disc looses its atomic gas content as widely discussed in the various HRS follow-up papers \citep{bos+14c,cor+16}.

At variance with the HI mass the BLF of the H$_2$ mass is relatively strong correlated with the \Kb~ luminosity (see Figure~\ref{MH2MF} and Table~\ref{tab:coeff}).
The correlation shown in Figure~\ref{MH2MF} reflects the relation between the stellar mass and the molecular gas mass within the sample with the scatter due to the HI-deficient galaxies 
\citep{bos+14b}. \citet{bos+14c} have used the $M(H_2)$ versus stellar mass, $M_{\rm star}$, scaling relation to define the H$_2$-deficiency parameter as the difference, on logarithmic scale, between the expected and observed molecular gas mass for a galaxy of given stellar mass. This molecular hydrogen deficiency is considered as a proxy for galaxy interactions with the surrounding cluster environment.
The molecular gas and the extension of the molecular disc are also affected by the presence of the cluster galaxies and on average these galaxies have a lower molecular
content than galaxies in the field.
A similar finding is reported by \citet{fum+09} who find that molecular deficient galaxies form stars at a lower rate or have dimmer far infrared fluxes
than gas rich galaxies, as expected if the star formation rate is determined by the molecular hydrogen content.
A different view has been proposed by \citet{mok+16} who argue that Virgo galaxies have longer molecular gas depletion times compared to group galaxies, due to their higher H$_2$ masses and lower star formation rates and suggest that the longer depletion times may be a result of heating processes in the cluster environment
or differences in the turbulent pressure. This issue requires further studies and is not settled yet.

Figure~\ref{MH2MF} displays the $\rm H_2$ MF derived from the BLF (Figure~\ref{MH2MF}) 
 compared with the predictions by \citet{lag+11}. At masses lower than $10^8 \rm M_\odot$ the HRS sample may miss galaxies with low content of molecular hydrogen.
However, very few samples in the Local Universe are complete in molecular hydrogen and the data of galaxies with very low molecular content in unbiased samples are still scanty 
\citep[i.e.][]{bot+16}. Previous molecular MFs of nearby galaxies have been derived from the CO MF. \citet{ker+03} used an incomplete CO sample based on a far-IR selection
and exploiting the correlation with the 60$\mu$m luminosity. The resulting CO MF is, therefore, biased towards gas rich galaxies. An updated estimate of
the $\rm H_2$ MF, based on an empirical and variable CO-H$_2$ conversion factor, was presented by \citet{obr+09}. We use in Figure~\ref{MH2MF} the molecular mass function derived from the $L^\prime(CO)$ luminosity distribution of \citet{san+17} from the COLD GASS (CO legacy data base for GASS; \citet{san+11}) survey. This last survey, although biased towards massive galaxies (stellar mass, $M_{\rm star} >$~a~few~$10^{9} M_\odot$, \citet{san+11,san+17}), i.e. it might not sample a sufficiently large dynamic range in $M_{\rm star}$ to trace a fair distribution, is at present the only survey with a large enough database to allow a fair reconstruction of the $L^\prime(CO)$ luminosity distribution. However, this sample too is not unbiased, i.e. it is not CO-selected.

The comparison shown in Figure~\ref{MH2MF} of the molecular mass function derived from the HRS and that from the COLD GASS sample is only indicative. In addition to the issues discussed above
we lack the information about the galaxy properties to apply the luminosity dependent conversion factor between $L^\prime(CO)$ and $M(H_2)$ equal to the one used by \citet{bos+14b}. 
What we show in Figure~\ref{MH2MF} is our derived $H_2$~MF using a constant conversion factor
($\rm \alpha_{\rm CO} = 3.6 M_\odot/(K km s^{-1} pc^2)$.
Moreover, the completeness at low molecular masses is for both samples very poor and below $log(M(H_2))=8.4$M$_\odot$ nothing can be inferred.

\section{Discussion}\label{discussion}

The fundamental goal for theoretical models of galaxy formation and evolution is reproducing the observed statistical distributions (such as LFs, stellar and cold gas MFs)
of the global properties of the galaxy populations at different cosmic epochs.

On the one hand, most of the models are not able to reconstruct the whole spectrum of data, commonly used to fix the parameters, and to predict the evolution at larger redshifts.
Because of the large uncertainties in the theory associated with the physics of the SF, stellar and AGN feedback and environmental effects, key is tuning the multi-parameter space by fitting the observed physical properties of the galaxies in the Local Universe. In addition, many free parameters are frequently degenerate with each other, and the tuned recipes make these models more or less successful in predicting the galaxy evolution over cosmic time.

On the other hand, from the observational side, the building up of samples sufficiently large to be statistically meaningful and with a wide wavelength range to cover the whole
spectrum of observed properties is laborious. But this would be the only way to allow a fair comparison with models and to keep biases and systematics fully under control.

We have discussed extensively the limitations of the HRS sample and constrained its biases and selection effects while discussing the individual mass and luminosity functions.
The sample is strongly limited in statistical significance by the small number of sources which does not allow to fully constrain the properties of the various functions. However, it is the only local sample which has a large coverage in wavelengths for which many physical properties can be simultaneously studied. 
The large number of observations and the original well defined selection in the \Kb~  have been used to define several LFs and MFs presented in this work which
can be used to constrain the galaxy formation models.

At a first order, the BLF and BMF that we estimate are fairly 
comparable to those derived in the literature given the wide variety of functions published not always 
consistent one another. Just for an example, the SFR LF seems to be the most different from those derived in the
literature.

It is clear that the SMF and the atomic gas MF, are very well and better determined in much larger local samples, but in the cases as the dust and the molecular MF, for which data are either scanty or not well constrained, the functions determined from the HRS show good quality and at the same level or even better than those found in the literature.

Furthermore, the HRS is composed of galaxies located in a wide range of environments, from the general field to the core of the Virgo cluster,
the largest concentration of galaxies in the nearby Universe. It is thus ideally defined to study in great detail environmental effects on the different 
galactic components (stars, gas, dust). Thanks to its proximity ($\sim$ 20 Mpc) and to the quality of the multifrequency data gathered so far, this sample
is a unique laboratory for studying the role of mass and environmental quenching and feedback on galaxy evolution down to sub kpc scales.

\subsection{The local LFs, the IR luminosity and the SFR densities}

The IR LF derived from the bivariate PDF $L_{\rm K}$-$L_{\rm IR}$ is well in agreement with that extracted from blind far-IR survey, it deviates mainly at low-IR luminosity where the HRS likely misses low luminosity galaxies.
The selection in the \Kb~, as discussed in \S ~\ref{LK-LIR}, may miss low surface density objects, faint optical galaxies and galaxies with IR luminosity larger than that expected for a given \Kb~ luminosity. Overall the agreement with the caveats mentioned above is good.

The present derivation of the luminosity functions allow us to derive the local extragalactic luminosity density. This latter is computed integrating the functional form of the
LFs within the limits where the function is defined $L(IR)=2\times10^8  \div 6 \times10^{10}  L_\odot$.
The  density of the IR luminosity in the local Universe, measured from the HRS IR LF shown in Figure~\ref{LIRLF} turns out to be 1.5~$10^7L_\odot$Mpc$^{-3}$,
a factor five lower than that reported in \citet{mar+16}, 8.3~$10^7L_\odot$Mpc$^{-3}$. The HRS misses starburst galaxies because of the small sampled volume.

The SFR function derived from the bivariate PDF $L_K$-${\rm SFR}$ as discussed in \S ~\ref{LK-SFR} shows a very different behaviour from those
derived from $H\alpha$ surveys and from blind IR surveys mainly at large values of the SFRs. 
This reflects two problems. First, the large difference in sampling the local SFR from optical and IR samples and secondly  the inference of the SFR from the observables with the optical values which are largely affected by the uncertainties on the attenuation correction factors. These latter depend on parameters such as stellar
mass and dust temperature, and our poor understanding of the relation between the IRX ratio (L(IR)/L(UV)) and the UV spectral slope \citep[see for extensive discussion, ][]{wan+16}.

Likely because the HRS SFR function misses large values of the SFR, due to the lack of starburst galaxies, the local SFR density computed on this sample
is a factor of two below that determined from other local surveys.
The SFR density is inferred integrating the derived SFR function shown in Figure~\ref{SFRLF} (within the integration limits 
$SFR= 0.01 \div 15 M_\odot/year$) and turns out to be $(1.6\pm0.4) 10^{-3} M_\odot$ yr$^{-1}$ Mpc$^{-3}$ which is a factor 2 lower than that derived from other optical \citep{gun+13} and 5-10 times lower than that derived from IR suverys \citep[see][and reference therein]{cle+13,mar+16}.
This is not surprising as a large scatter in the local SFRD estimates using different SFR diagnostics is seen. In addition galaxies in the Virgo cluster show a reduced SFR.
The $H\alpha$ measurements present the largest scatter among different published results \citep[see Figure 11 in][]{bos+15}, \citep[][and reference therein]{mar+16}.

\subsection{The local mass functions and local mass densities}

 As discussed in section ~\ref{LK-M*} the SMF of the HRS shows a deficit of small galaxies due to the limit in the original selection in the \Kb ~~and to the poor sampling of low surface brightness galaxies.
The computed local stellar mass density of the HRS (integrating the functional form over the range $M_{stellar}=10^9  \div 2 \times10^{11}  M_\odot$) turns out to be 2.25~$10^8M_\odot$ Mpc$^{-3}$, within a factor of 2 from that computed integrating the best fit of the SMF given by \citet{bal+12}.

The dust mass function of the HRS sample follows closely the same behaviour as those derived from other blind IR surveys. The large scatter shown among the different functions reflect the uncertainties related to the physical and chemical properties of the dust grains. The derived local dust mass density has a value
consequently in between the value derived with the \citet{dun+11,bee+17}'s and \citet{cle+13}'s mass functions.
The dust mass local density $\sim 8~10^{4} M_\odot$ Mpc$^{-3}$, obtained integrating the functional form (Eq.~\ref{SB}) over the range $10^5\div5~10^8 M_\odot$, agrees within the uncertainties with those derived from the other DMFs computed so far and rescaled by \citet{bee+17} at the same value of the dust absorption coeffecient, $\sim 1.5~10^{5} M_\odot$ Mpc$^{-3}$.
\citet{dri+18} report from their analysis on the GAMA survey an average value of the local dust mass density of $\sim 1.4~10^{5} M_\odot$ Mpc$^{-3}$.

Figure~\ref{MHIMF} shows the atomic gas function and highlights that $HI$-MF derived from the HRS data differs substantially at values $M(HI)>$ a few $\times10^9M_\odot$. This is due to the very weak correlation between the HI mass with the \Kb~ luminosity  (see Figure~\ref{MHIMF} and Table~\ref{tab:coeff})
which does not allow to strongly constrain the bivariate and as a consequence the construction of the atomic gas mass function is poorly determined. 
In particular we find a deficiency of large mass objects in the HRS survey because its selection in the \Kb~ misses galaxies with large values of the atomic gas found
in HI dedicated blind surveys \citep{zwa+05,mar+10,hop+15,jon+18}.
Moreover, the HRS contains more HI deficient objects than normal field galaxies, due to the presence of the Virgo cluster, and the likely direct stripping of ISM from the disc (e.g., ram pressure) \citep{cor+16,bos+14c}.

\begin{table*}
\label{densities}
\caption{local luminosity and mass densities}                 
\centering          
\begin{tabular}{c c}     
\hline\hline       
\\
Luminosity/Mass & Local density value \\

\hline    
\\
IR luminosity & 1.5~$10^7L_\odot$Mpc$^{-3}$\\
SFR & $1.6 \times 10^{-3}  M_\odot$ yr$^{-1}$ Mpc$^{-3}$ \\     
Stellar mass  &  2.25~$10^8M_\odot$ Mpc$^{-3}$\\
Dust mass & $8~10^4 M_\odot$ Mpc$^{-3}$\\
Molecular mass  & $10^{7}M_\odot $Mpc$^{-3}$\\

\\
\hline           
    
\end{tabular}
\end{table*}

The molecular mass function reported in Figure~\ref{MH2MF} is the first function built on a complete sample, although the completeness is in the \Kb~.
The $H_2$ mass is relatively strong correlated with the \Kb~ luminosity due to relation between the stellar mass and the molecular gas mass within the sample with the scatter due to the HI-deficient galaxies.
The derived $H_2$ MF when compared with the predictions by \citet{lag+11}, shows a deficit at masses lower than $10^8M_\odot$ where the HRS sample may miss galaxies with low content of molecular hydrogen.

We have derived a very rough molecular mass function from the best-fit of the CO luminosity distribution by \citet{san+17} and compare this function to the one computed over the HRS sample. The comparision is only indicative. It shows overall good agreement but at small molecular mass where neither the HRS nor the COLD GASS sample are complete and therefore we cannot infer any meaningful conclusion. 
The molecular mass local density turns out to be $10^{7}M_\odot $Mpc$^{-3}$.

Table~\ref{densities} summarises the values of the local luminosity and mass density derived from our analysis.

\section{Conclusions}\label{conclusions}

The construction of the LFs and the MFs has made possible using the bivariate based on the \Kb~ selection. 
We have discussed the LFs and MFs derived from the HRS and compared with the same LFs and MFs derived from local samples selected in complete different ways.
This comparison highlights the limits and biases inherent to the HRS but also its strength as representative sample of the Local Universe.

The analysis shows that the behaviour of the morphological (optically selected) subsamples is quite different and a statistically meaningful result can be obtained over the whole HRS sample only from the relationship between the \Kb~ and the stellar mass. The same analysis with the other physical quantities (dust and gas masses, far-IR luminosity and star formation rate) has to be restricted to the late-type galaxy (LTG) subsample.
The LFs, MFs of LTGs are generally dependent on the \Kb~ and the various dependencies are discussed in detail. We are able to derive the corresponding LFs and MFs and compare with those computed with other samples and with results from galaxy formation simulations.

The analysis reported in this paper represents a fundamental local benchmark to compare with models of 
galaxy evolution. The HRS is designed to provide a concise view of the large galaxies in our local Universe. 
The results found in this work could therefore be representative for late type galaxies. The derived relations
can be applied to a larger set of local galaxies and can be compared with a similar analysis at higher redshift.

\begin{acknowledgements}
This research has made use of data from the HRS project. HRS is a Herschel Key Programme utilising guaranteed time from the 
SPIRE instrument team, ESAC scientists and a mission scientist. The HRS data was accessed
through the Herschel Database in Marseille (HeDaM - http://hedam.lam.fr) 
operated by CeSAM and hosted by the Laboratoire d’Astrophysique de Marseille.
We acknowledge financial support from Programme National de Cosmologie and Galaxies (PNCG) 
of CNRS/INSU, France.
Parts of this research were conducted by the Australian Research Council Centre of Excellence for All Sky Astrophysics in 3 Dimensions (ASTRO 3D), through project number CE170100013. 

P.A. warmly thanks Ken Tatematsu-san for his kind invitation to Japan and his hospitality at the Nobeyama Observatory. She warmly thanks NAOJ-Chile for her stay at Mitaka where this work has been completed.

We thank an anonynous referee whose help has largely improved the readability of the paper

\end{acknowledgements}

\clearpage

 \begin{figure*}[h]
   \centering{}
       \resizebox{\vsize}{!}{\includegraphics{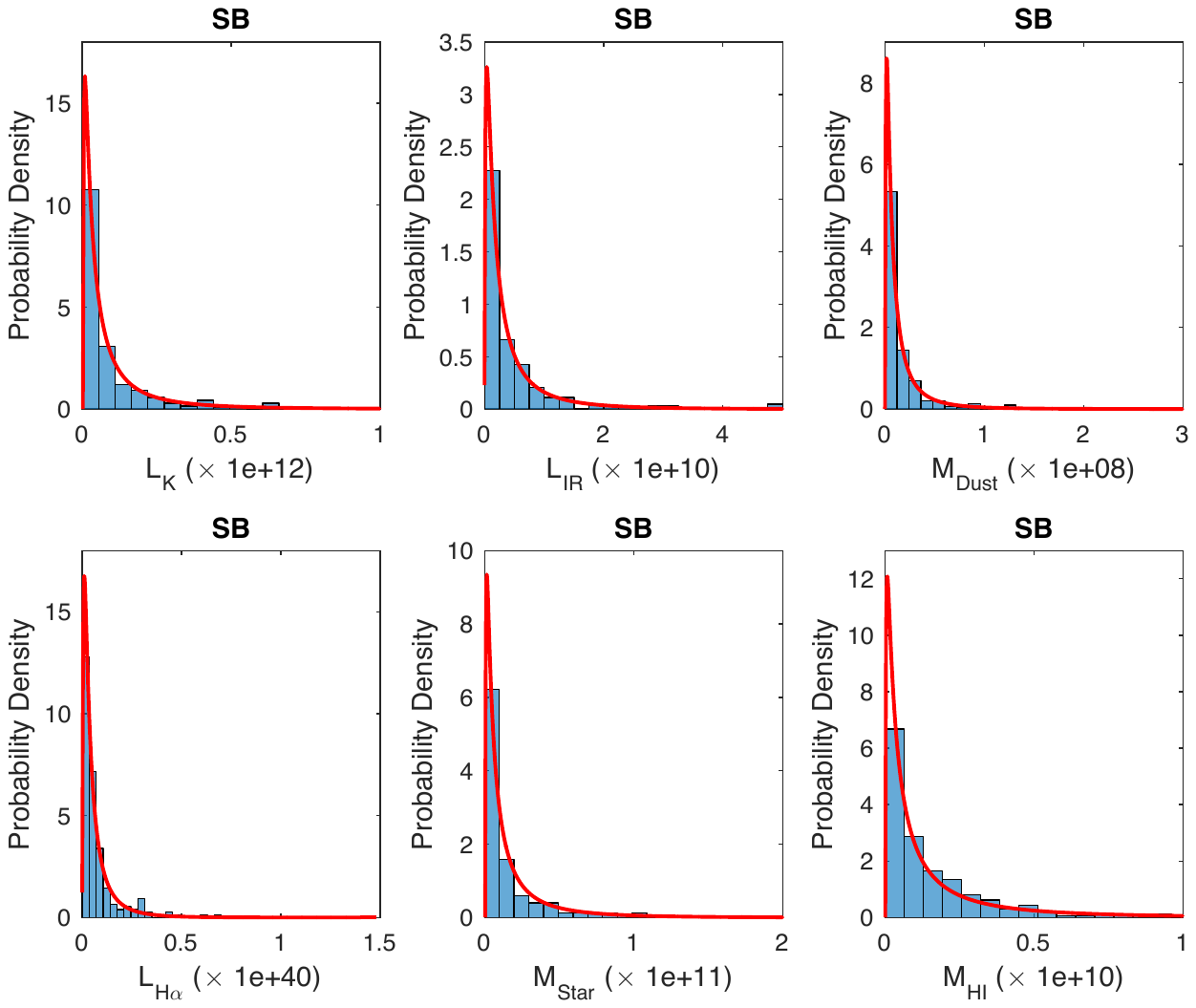}}
        \caption{Histograms of the data in table~\ref{tabledata} and corresponding fitted Johnson PDF. In all cases the SB family has been selected.
        The best fit parameters of the equation~\ref{SB} are reported on Table~\ref{tab:coeff}. These PDFs have been used to estimate the bivariate PDF shown in Figures~\eqref{LIRLF}-\eqref{SFRLF}.}
        \label{fig:histograms}
    \end{figure*}

\begin{landscape}
  \begin{figure}
 \centering{}
       \resizebox{\hsize}{!}{\includegraphics{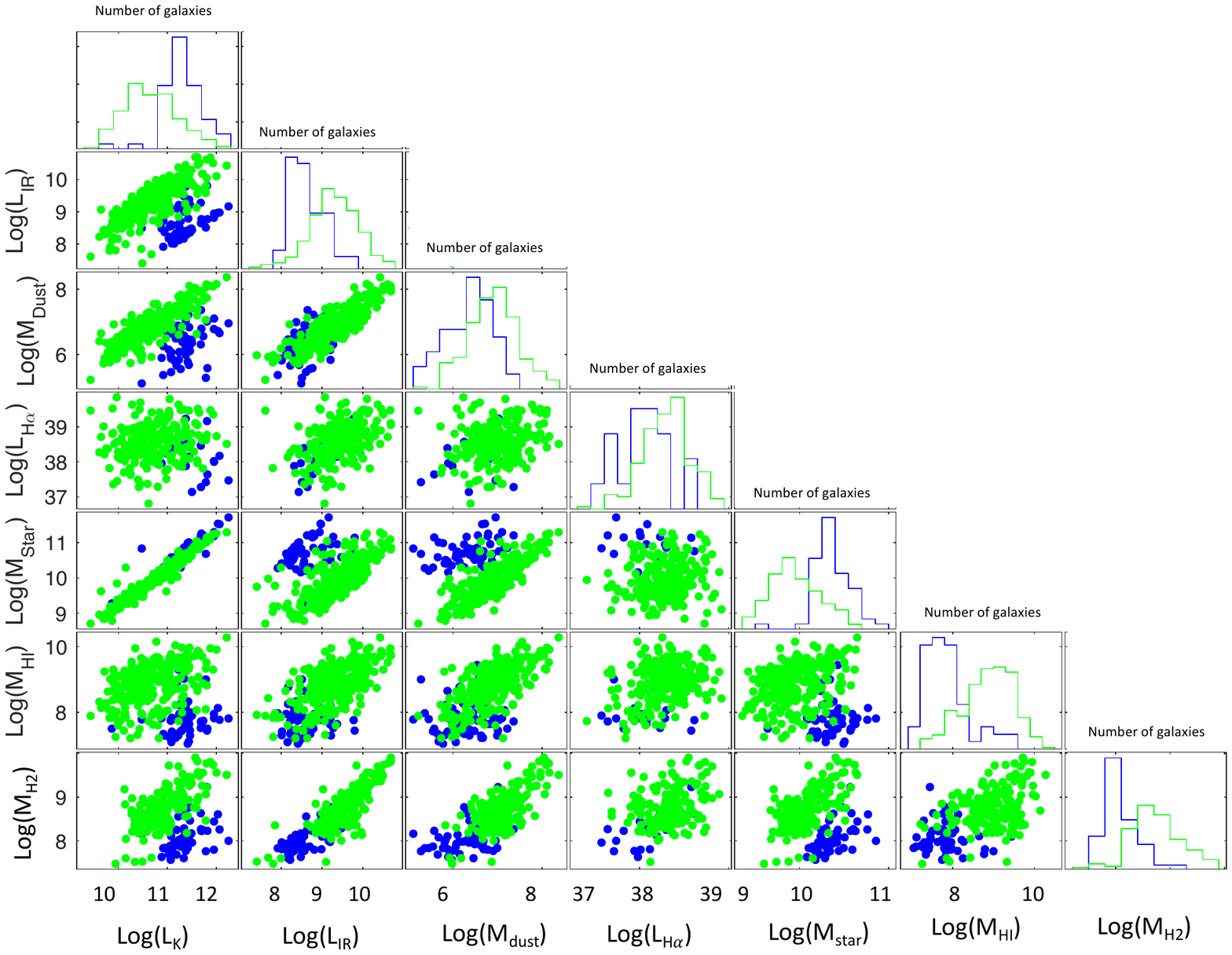}}
        \caption{Scatterplots and histograms for the data. Blue dots correspond to the objects with morphological type $< 3$ (ET ), green dots to objects with morphological type $\geq 3$ (LT ). Because of the lack of data at some wavelengths the various panels contain a different number of points.}
        \label{correlations}
    \end{figure}
\end{landscape}

  \begin{figure*}
       \resizebox{\hsize}{!}{\includegraphics{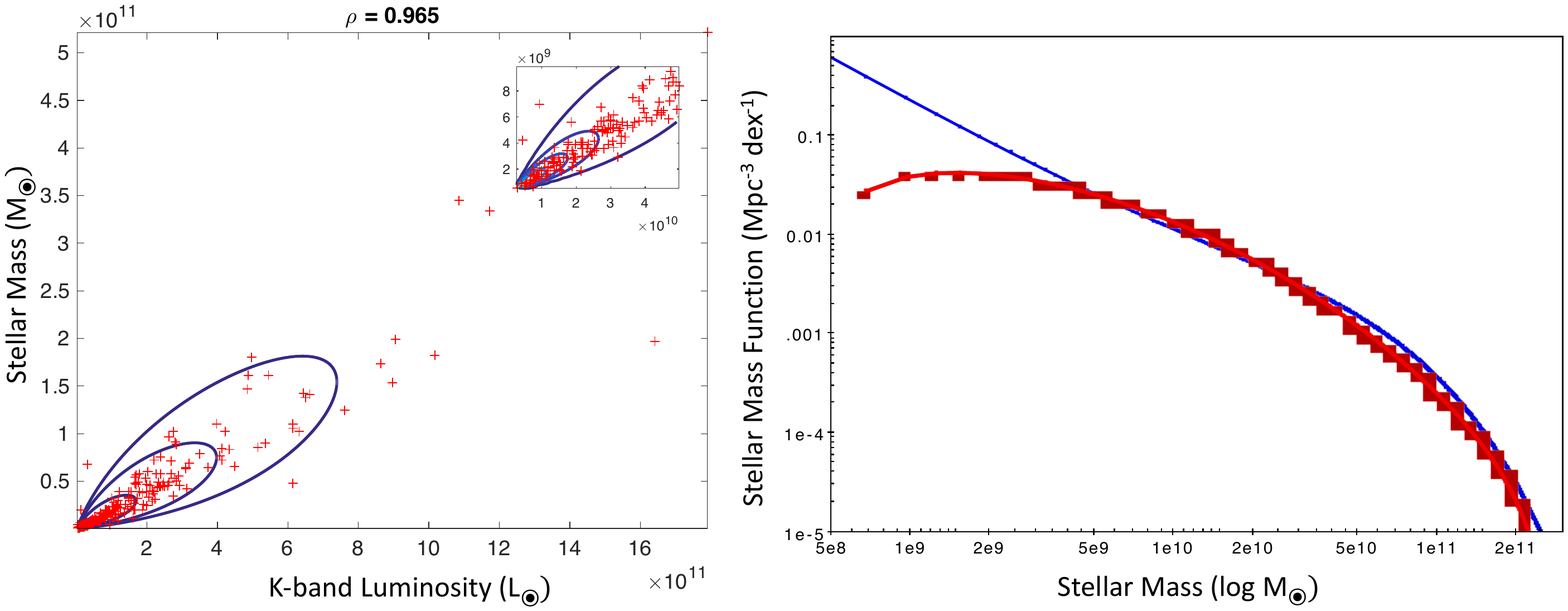}}
        \caption{(a) Bivariate PDF ${L_K}$-${M_{\rm star}}$: estimated bivariate PDF shown in linear  scale (and enlarged in the inset) for the \Kb~ band with the stellar masses.
Contour lines correspond to the levels 0.1, 0.3, 0.5, 0.7, 0.9. These values correspond to the fraction of the peak value of the BLF  that is set to one. (b) Reconstructed stellar mass function of the HRS sample (red line) shown together with the best fit of the galactic mass function given by \citet{bal+12} for the GAMA survey (blue line). 
}
        \label{SMF}
    \end{figure*}
  \begin{figure*}
       \resizebox{\hsize}{!}{\includegraphics{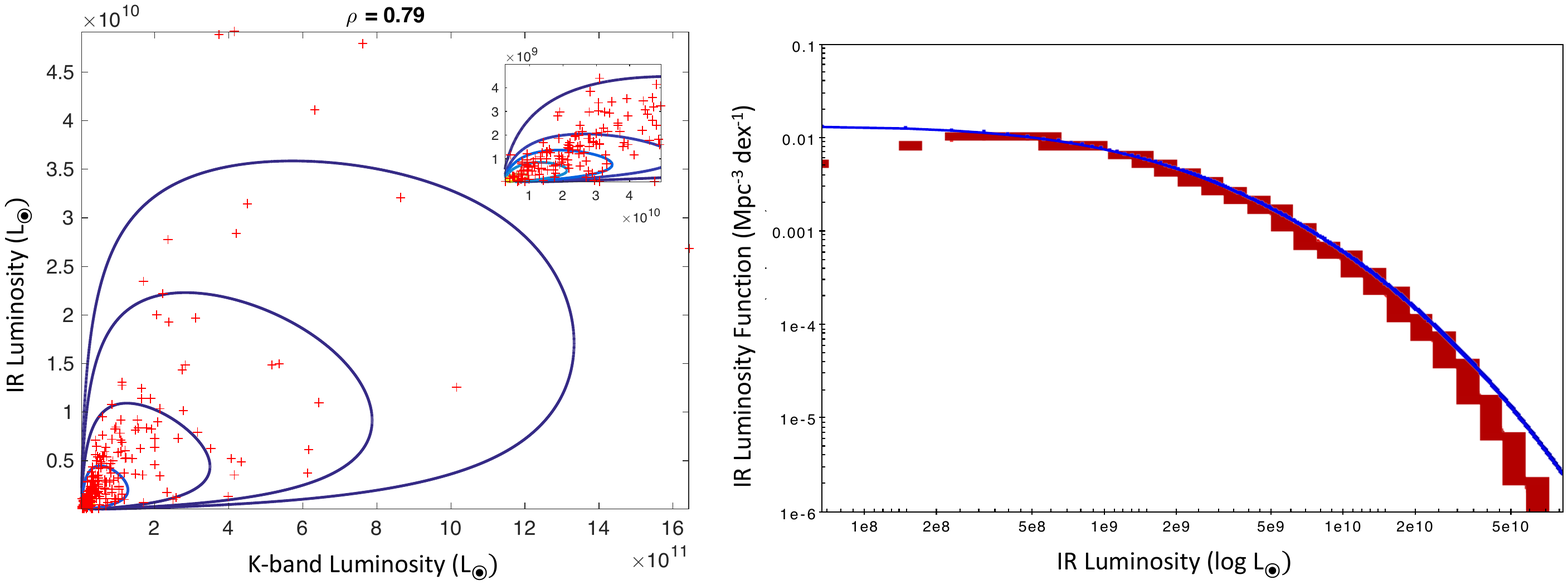}}
        \caption{(a) Bivariate PDF ${L(K)}$-${L(IR)}$ for LTGs only: estimated bivariate PDF shown in linear  scale (and enlarged in the inset) for the \Kb~ with the infrared luminosities.
Contour lines correspond to the levels 0.1, 0.3, 0.5, 0.7, 0.9. These values correspond to the fraction of the peak value of the BLF  that is set to one. (b) Reconstructed IR luminosity function of the HRS sample (red line) shown together with the LF computed over local blind IR surveys (blue line) by \citet{mar+16}. Errobars are estimated through a bootstrapping technique.}\label{LIRLF}
    \end{figure*}

  \begin{figure*}
       \resizebox{\hsize}{!}{\includegraphics{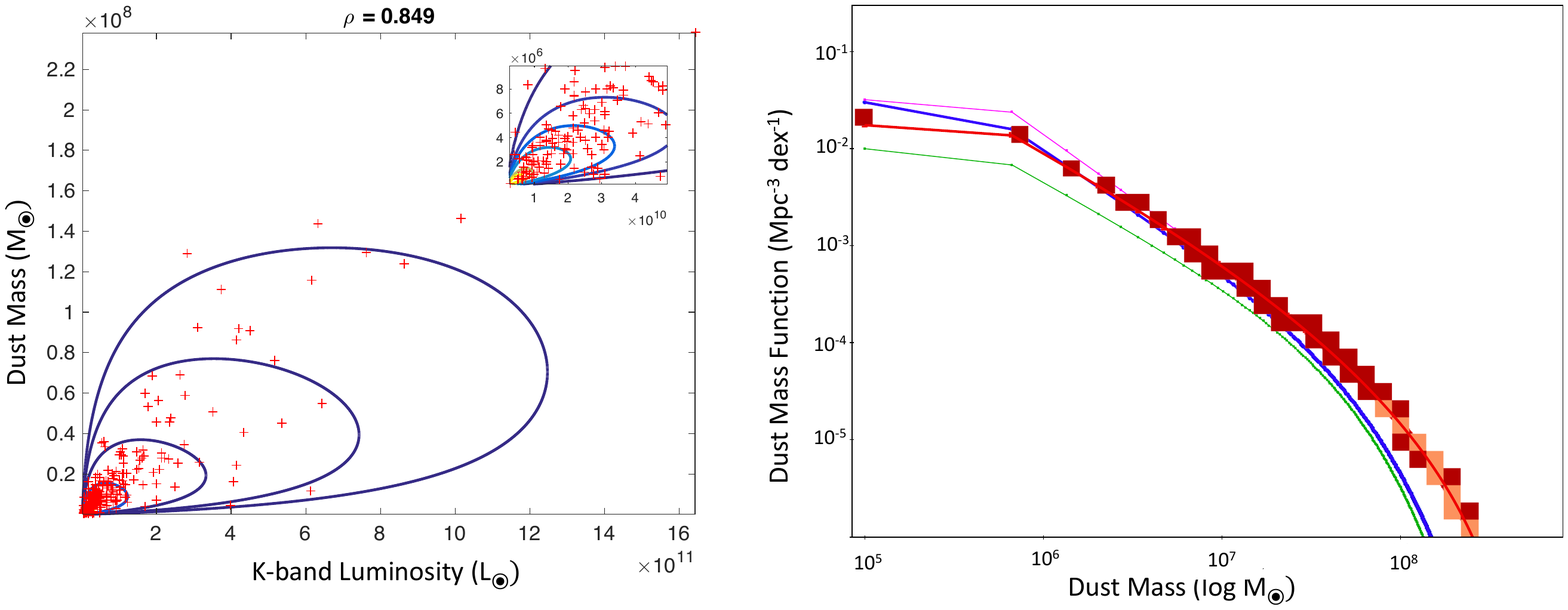}}
        \caption{(a) Bivariate PDF ${L(K)}$-${M_{\rm dust}}$, same as Figure~\ref{LIRLF} for the \Kb~ luminosity and the dust mass. (b) Reconstructed Dust Mass Function of the HRS sample (red line) shown together with the DMFs computed over local blind IR surveys (\citep{dun+11}, green, \citep{bee+17}, blue, \citep{cle+13} purple and grey).}
        \label{MDF}
    \end{figure*}

 \begin{figure*}
       \resizebox{\hsize}{!}{\includegraphics{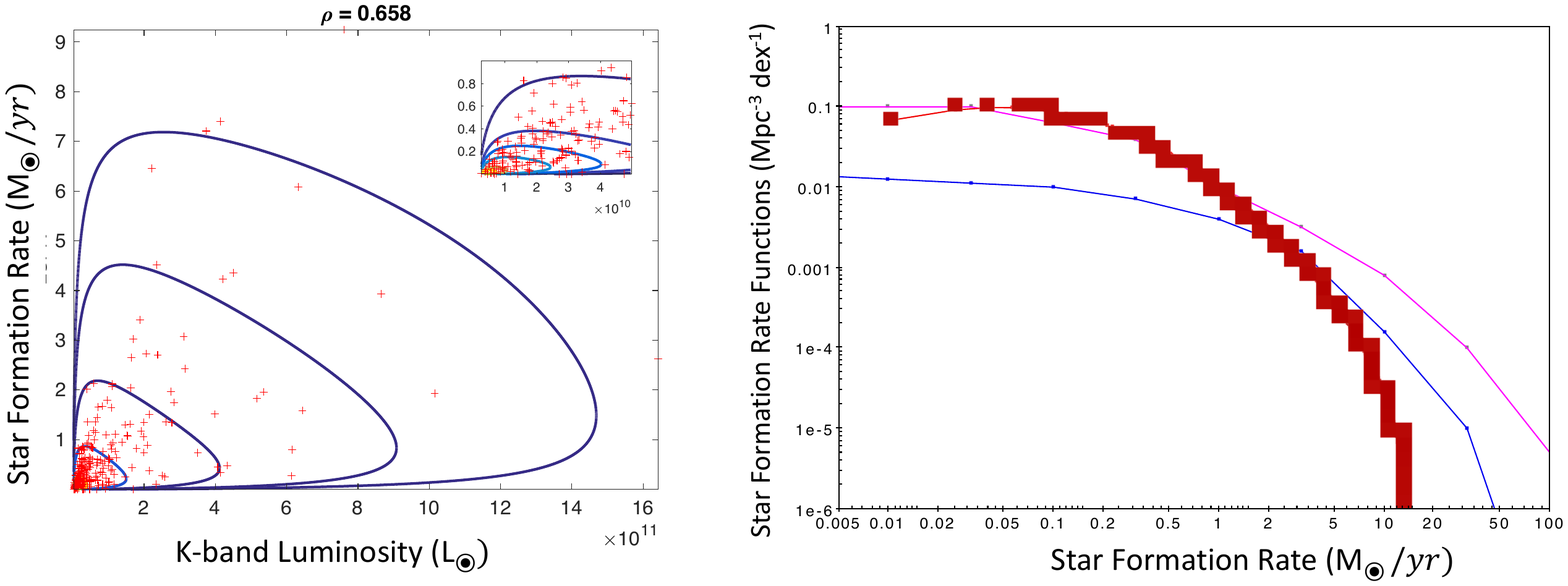}}
        \caption{(a) Bivariate PDF  ${L(K)}$-${\rm SFR}$, same as Figure~\ref{SMF} for the \Kb~ luminosity and the SFR. (b) Reconstructed SFR LF for the HRS sample (red) compared with that of \citet{gun+13} (blue) and \citet{cle+13} (purple).}
        \label{SFRLF}
    \end{figure*}    

   \begin{figure*}
       \resizebox{\hsize}{!}{\includegraphics{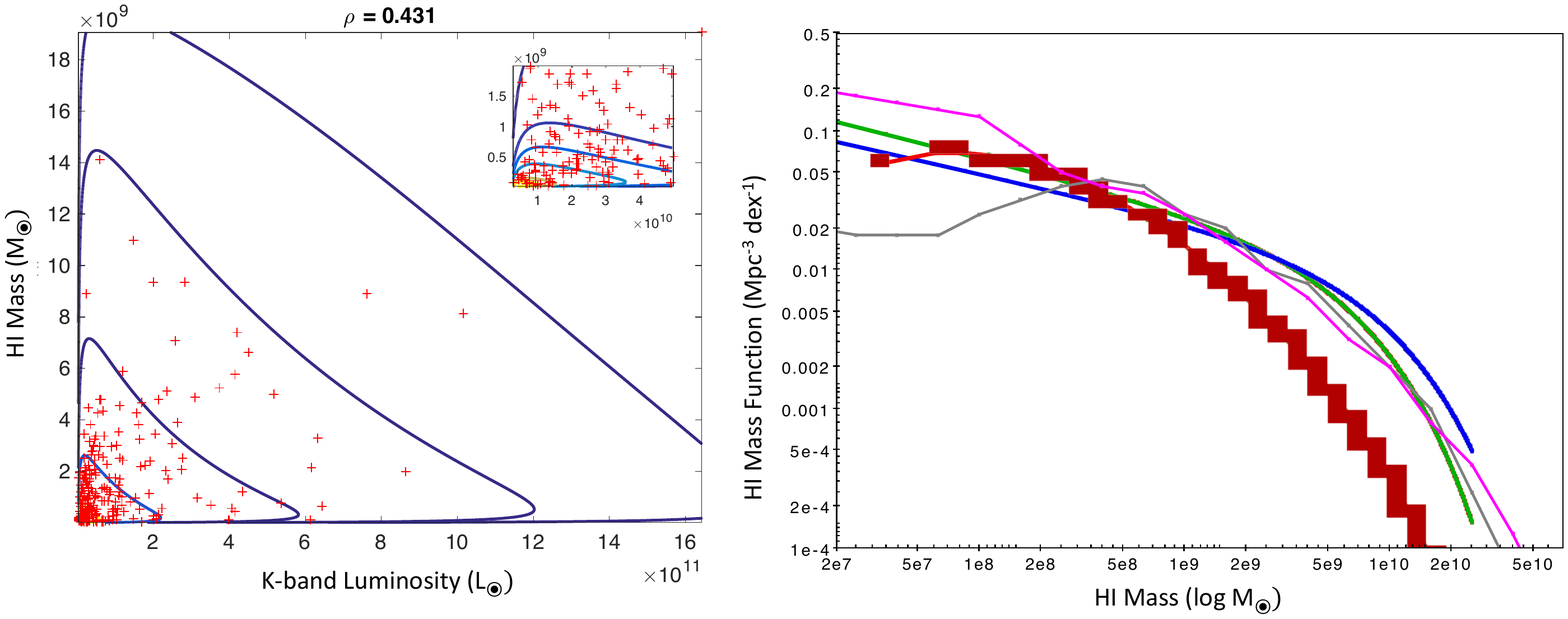}}
        \caption{(a) Bivariate PDF  ${\L(K)}$-${M(HI)}$, same as Figure~\ref{LIRLF} for the \Kb~ luminosity and the atomic gas mass. (b) Reconstructed atomic gas mass function of the HRS sample (red curve) shown together with the gas mass computed in the local Universe by \citep{mar+10,zwa+05} (blue and green curves respectively) and models \citep{pop+14,lag+11} (purple and grey points/lines)}
        \label{MHIMF} 
    \end{figure*}

   \begin{figure*}
       \resizebox{\hsize}{!}{\includegraphics{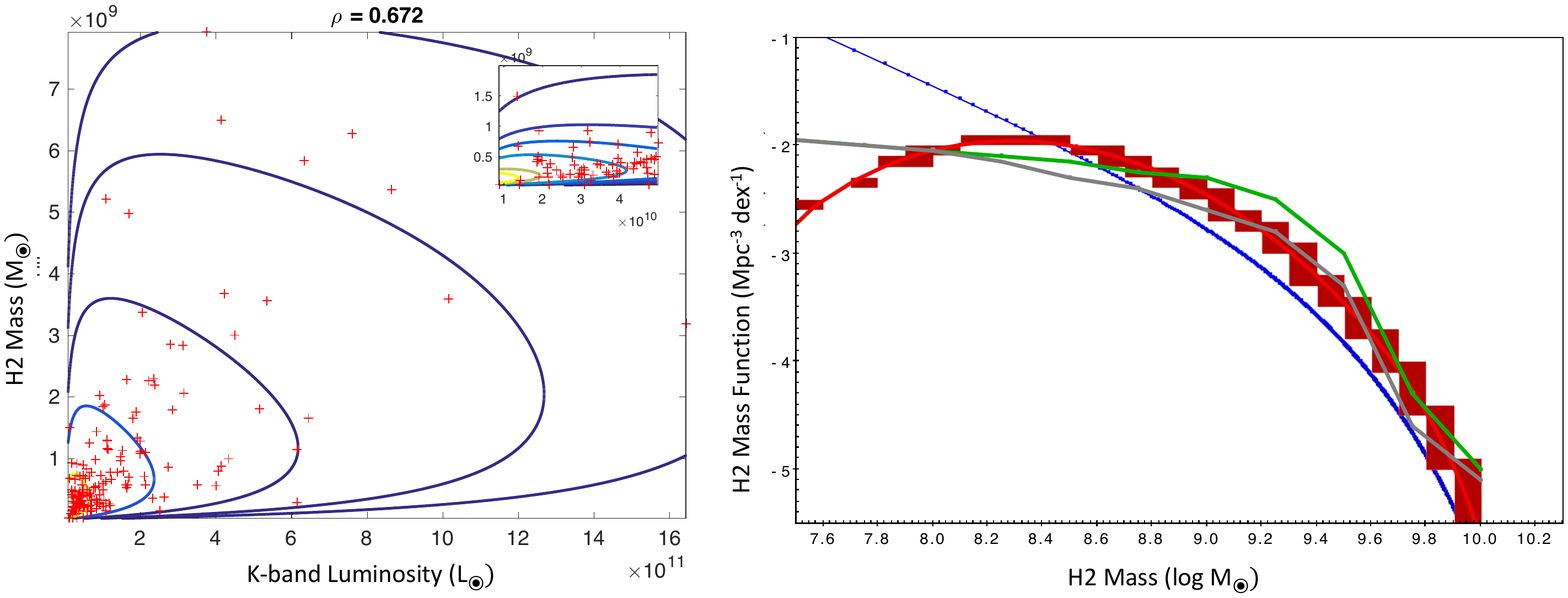}}
        \caption{(a) Bivariate PDF  ${L(K)}$-${M(H_2)}$, same as Figure~\ref{LIRLF} for the \Kb~ luminosity and the molecular gas mass. (b) Reconstructed molecular gas mass function of the HRS sample shown together with the predicted ones by \citet{lag+11} (green and grey lines), and the molecular mass function derived from the $\rm L^\prime (CO)$ function from \citet{san+17} (blue line).}
        \label{MH2MF}
    \end{figure*}

\end{document}